\documentclass[12pt]{article}
\usepackage{amsmath,amssymb,amsthm,amsxtra,overpic,bbm,bm,epsfig,subfigure}
\usepackage{hyperref}
\usepackage{mathrsfs}
\usepackage{graphicx}
\usepackage{multirow}
\usepackage{array}
\usepackage{color}
\usepackage{booktabs}
\usepackage{subfigure}
\usepackage{graphicx}
\usepackage{float}
\usepackage{hyperref}
\hypersetup{citecolor=blue}
\usepackage{array}
\usepackage{colordvi}
\usepackage{mathtools}
\usepackage{multirow}
\usepackage{color}
\usepackage{comment}
\usepackage{float}
\usepackage{cite}
\usepackage{slashed,stmaryrd}
\usepackage{stackengine} % adjust row height in table

\newcommand\xrowht[2][0]{\addstackgap[.5\dimexpr#2\relax]{\vphantom{#1}}}

\begin{document}

\vspace{0.2cm}

\begin{center}
{\Large\bf Inverse Seesaw Model with a Modular $S^{}_4$ Symmetry:\\ Lepton Flavor Mixing and Warm Dark Matter}
\end{center}

\vspace{0.2cm}

\begin{center}
{\bf Xinyi Zhang}~$^{a}$~\footnote{Email: zhangxinyi@ihep.ac.cn},
\quad
{\bf Shun Zhou}~$^{a,~b}$~\footnote{Email: zhoush@ihep.ac.cn (corresponding author)}
\\
\vspace{0.2cm}
{\small $^a$Institute of High Energy Physics, Chinese Academy of Sciences, Beijing 100049, China\\
$^b$School of Physical Sciences, University of Chinese Academy of Sciences, Beijing 100049, China}
\end{center}

\vspace{1.5cm}

\begin{abstract}
In this paper, we present a systematic investigation on simple inverse seesaw models for neutrino masses and flavor mixing based on the modular $S^{}_4$ symmetry. Two right-handed neutrinos and three extra fermion singlets are introduced to account for light neutrino masses through the inverse seesaw mechanism, and to provide a keV-mass sterile neutrino as the candidate for warm dark matter in our Universe. Considering all possible modular forms with weights no larger than four, we obtain twelve models, among which we find one is in excellent agreement with the observed lepton mass spectra and flavor mixing. Moreover, we explore the allowed range of the sterile neutrino mass and mixing angles, by taking into account the direct search of $X$-ray line and the Lyman-$\alpha$ observations. The model predictions for neutrino mixing parameters and the dark matter abundance will be readily testable in future neutrino oscillation experiments and cosmological observations.
\end{abstract}

\newpage

\def\thefootnote{\arabic{footnote}}
\setcounter{footnote}{0}

\section{Introduction}

Tiny but nonzero neutrino masses, as indicated by successful neutrino oscillation experiments in the past few decades~\cite{Kajita:2016cak, McDonald:2016ixn}, call for new physics beyond the standard model of particle physics (SM). On the other hand, cosmological observations reveal that dark matter exists in our Universe, but none of the SM particles could be the primary candidate for dark matter~\cite{Bergstrom:2000pn, Bertone:2004pz, Feng:2010gw, Bertone:2016nfn}. After decades of exploration, the origin of neutrino masses and the basic properties of dark matter remain mysterious, constantly attracting tremendous attention.

As for neutrino masses, the simplest explanation may be to extend the SM by introducing three right-handed neutrinos and implement the seesaw mechanism~\cite{Minkowski:1977sc,Mohapatra:1979ia,GellMann:1980vs}. In this canonical type-I seesaw model, the effective mass matrix of three ordinary neutrinos is given by $M^{}_\nu \approx - M^{}_{\rm D} M^{-1}_{N} M^{\rm T}_{\rm D}$, where the Dirac neutrino mass matrix $M^{}_{\rm D}$ comes out from the spontaneous breakdown of electroweak gauge symmetry and thus is naturally around the electroweak scale $\Lambda^{}_{\rm EW} = 10^{2}~{\rm GeV}$. Therefore, the Majorana mass matrix $M^{}_N$ of right-handed neutrinos is expected to be around $\Lambda^{}_{N} = 10^{14}~{\rm GeV}$ to generate the sub-eV masses of three light neutrinos. Unfortunately, such a high-energy scale $\Lambda^{}_{N} = 10^{14}~{\rm GeV}$ renders the type-I seesaw model impossible to be tested directly in the terrestrial collider experiments.

The inverse seesaw (ISS) mechanism~\cite{Mohapatra:1986bd, GonzalezGarcia:1988rw, Deppisch:2004fa} offers a possible way out of this testability problem. In the most general ISS model, there are three left-handed fermion singlets $S^{}_{i{\rm L}}$ (for $i = 1, 2, 3$), three right-handed neutrinos $N^{}_{i{\rm R}}$ (for $i = 1, 2, 3$), and one scalar singlet $\Phi$, in addition to the SM particles. The gauge-invariant Lagrangian for neutrino masses can be written as
\begin{eqnarray}
-{\cal L}^{}_{\nu} =  \overline{\ell^{}_{\rm L}} Y^{}_\nu \widetilde{H} N^{}_{\rm R} + \overline{S^{}_{\rm L}} Y^{}_\mathrm{S} N^{}_{\rm R} \Phi + \frac{1}{2} \overline{S^{}_{\rm L}} \mu S^{\rm C}_{\rm L} + {\rm h.c.}\; ,
\label{eq:lag}
%     (1)
\end{eqnarray}
where $\ell^{}_{\rm L}$ and $\widetilde{H}\equiv i\sigma^2 H^*$ are left-handed lepton and Higgs doublets, $Y^{}_\nu$ and $Y^{}_\mathrm{S}$ are $3\times 3$ Yukawa coupling matrices, and $\mu$ is the Majorana mass matrix of $S^{}_{\rm L}$. After the Higgs doublet and scalar singlet acquire their vacuum expectation values (vev's), i.e., $\langle H \rangle$ and $\langle \Phi \rangle$, the gauge symmetry is spontaneously broken and we obtain the overall $9\times 9$ neutrino mass matrix
\begin{eqnarray}
{\cal M} = \left( \begin{matrix} {\bf 0} & M^{}_{\rm D} & {\bf 0} \cr M^{\rm T}_{\rm D} & {\bf 0} & M^{\rm T}_\mathrm{S} \cr {\bf 0} & M^{}_\mathrm{S} & \mu \end{matrix} \right) \; ,
\label{eq:M99}
%     (2)
\end{eqnarray}
where $M^{}_{\rm D} = Y^{}_\nu \langle H \rangle$ and $M^{}_\mathrm{S} = Y^{}_\mathrm{S} \langle \Phi \rangle$. The effective mass matrix of three ordinary neutrinos can be derived by diagonalizing the mass matrix in Eq.~(\ref{eq:M99}), namely,
\begin{eqnarray}
M^{}_\nu \approx - M^{}_{\rm D} M^{-1}_\mathrm{S}  \mu  \left(M^{}_{\rm D} M^{-1}_\mathrm{S} \right)^{\rm T} \; ,
\label{eq:issf}
%     (3) \cdot
\end{eqnarray}
where ${\cal O}(\mu) \ll {\cal O}(M^{}_{\rm D}) \ll {\cal O}(M^{}_\mathrm{S})$ has been assumed. It is straightforward to verify that ${\cal O}(M^{}_\nu) \sim 0.1~{\rm eV}$ can be realized by setting ${\cal O}(\mu) \sim 1 ~ {\rm keV}$ and ${\cal O}(M^{}_{\rm D}M^{-1}_\mathrm{S}) \sim 10^{-2}$. Two salient features of the ISS model can be observed. First, compared to the electroweak scale $\Lambda^{}_{\rm EW} = 10^2~{\rm GeV}$, the mass scale ${\cal O}(\mu) \sim 1~{\rm keV}$ is highly suppressed. This is reasonable according to 't Hooft's naturalness criterion~\cite{tHooft:1979rat}, since the Lagrangian in Eq.~(\ref{eq:lag}) gains a global $U(1)$ symmetry corresponding to the lepton number if $\mu$ is vanishing. Similar to the SM leptons, both $N^{}_\mathrm{R}$ and $S^{}_\mathrm{L}$ can be assigned with the lepton number $L = + 1$. Therefore, only the Majorana mass term of the fermion singlets, i.e., the $\mu$ term, violates the lepton number by two units. In this sense, the lightness of three ordinary neutrinos can be ascribed to the smallness of lepton number violation. However, the lepton-number-violating mass terms associated with $\overline{N^{\rm C}_{\rm R}} N^{}_{\rm R}$ and $\overline{\ell^{}_{\rm L}} \widetilde{H}  S^{\rm C}_{\rm L}$ are in general allowed as well. In the literature, the former case is also considered in the inverse seesaw framework, whereas the latter refers to the linear seesaw scenario~\cite{Malinsky:2005bi}. All these possibilities share two common features: (i) the lepton-number-violating terms are naturally small; (ii) the smallness of these terms suppresses the masses of ordinary neutrinos. For clarity, these lepton-number-violating mass terms are usually considered separately, as we shall do in the present work. Second, given ${\cal O}(M^{}_{\rm D}) \sim \Lambda^{}_{\rm EW} = 10^2~{\rm GeV}$, we immediately obtain ${\cal O}(M^{}_\mathrm{S}) \sim 10~{\rm TeV}$, implying that the mass scale of all the singlet fermions is well accessible to the large hadron colliders~\cite{Deppisch:2015qwa}.

In the present work, we construct a minimal but viable ISS model for neutrino masses, and further incorporate the modular $S^{}_4$ symmetry into the model to explain lepton flavor mixing. The motivation for such an investigation is two-fold. First, it is interesting to interpret tiny neutrino masses and provide a suitable candidate for dark matter at the same time. The ISS model with two right-handed neutrino singlets $N^{}_{i{\rm R}}$ (for $i = 1, 2$) and three left-handed fermion singlets $S^{}_{i{\rm L}}$ (for $i = 1, 2, 3$), which will be denoted as ISS(2,3)~\cite{Malinsky:2009df, Abada:2014vea}, serves as a perfect framework to achieve this goal. Second, although the ISS mechanism could account for tiny neutrino masses in a natural way, the flavor structures of lepton mass matrices remain unknown. Flavor symmetry has been a powerful tool in describing lepton flavor mixing. For recent reviews, see e.g., Refs.~\cite{Petcov:2017ggy,King:2017guk,Xing:2019vks}. The lepton flavor models based on discrete flavor symmetries suffer from the problems of too many new scalar fields (called ``flavons") and the flavons' vacuum alignments. For this reason, we implement the modular $S^{}_4$ symmetry~\cite{Feruglio:2017spp}, where no flavons are involved and the only source of symmetry breaking is the vev of the modulus $\tau$. Moreover, $N^{}_{i{\rm R}}$ (for $i = 1, 2$) and $S^{}_{i{\rm L}}$ (for $i = 1, 2, 3$) fit perfectly into the two- and three-dimensional irreducible representations of the $S^{}_4$ group. In the literature, there are a lot of studies on the $S^{}_4$ modular symmetry~\cite{Penedo:2018nmg, Novichkov:2018ovf, Novichkov:2019sqv, Okada:2019lzv, King:2019vhv, Criado:2019tzk, Wang:2019ovr, Gui-JunDing:2019wap} and those on ISS models (for an incomplete list, see Refs.~\cite{Malinsky:2009df, Hirsch:2009mx, Dev:2009aw, Abada:2014kba, CentellesChulia:2020dfh, Camara:2020efq}), but none on both. Refs.~\cite{Nomura:2019xsb,Nomura:2020cog} consider the ISS model with a modular $A^{}_4$ symmetry, but three pairs of fermion singlets are introduced and thus no dark matter candidate exists.
%in the former, while  the ${\rm SU}(2)$ multiplets are introduced to realize the TeV energy scale in the latter.

The remaining part of this paper is organized as follows. In Sec.~\ref{sec:model}, a simple ISS model with the modular $S^{}_4$ symmetry is presented. The model predictions are confronted with the global-fit results of neutrino oscillation data in Sec.~\ref{sec:ph}, and the allowed regions of model parameters are obtained. Sec.~\ref{sec:dm} is devoted to the discussions about the keV-mass sterile neutrino in our model and the observational constraints from dark matter abundance, $X$-ray line searches and cosmological structure formation. We summarize our main conclusions in Sec.~\ref{sec:conclusion}. The introduction to modular symmetries, the basics of the $S^{}_4$ group and the block diagonalization of neutrino mass matrix are given in Appendix~\ref{apd:modular}, \ref{apd:S4} and \ref{apd:block}, respectively.

\section{ISS Models with Modular $S^{}_4$ Symmetry}\label{sec:model}

One of the guiding principles for model building is simplicity. As first shown in Ref.~\cite{Malinsky:2009df}, two pairs of right-handed neutrino singlets $N^{}_{i{\rm R}}$ (for $i = 1, 2$) and left-handed fermion singlets $S^{}_{i{\rm L}}$ (for $i = 1, 2$) are sufficient and necessary to accommodate the observed neutrino mass-squared differences and lepton flavor mixing in the ISS model. In this section, we extend this minimal ISS model by an extra fermion singlet, namely, the ISS(2, 3) model. One benefit from such an extension is to provide a keV-mass sterile neutrino, which serves as the dark matter candidate. Another one is the appealing assignment of all these fermion singlets into the irreducible representations of the modular $S^{}_4$ group. More explicitly, we arrange two right-handed neutrinos $N^{}_{i{\rm R}}$ (for $i = 1, 2$) in the two-dimensional irreducible representation ${\bf 2}$ of $S^{}_4$, while three singlet fermions $S^{}_{i{\rm L}}$ (for $i = 1, 2, 3$) in the three-dimensional irreducible representation $\bf{3}$. The lepton doublets are assigned into the three-dimensional irreducible representation ${\bf{3}}$, and the right-handed charged leptons are arranged in one-dimensional irreducible representations $\bf{1}$ or $\bf{1}^\prime$.

Although it is in principle free to choose the weights of the modular forms for the Yukawa couplings, we perform a systematic study on models with modular-form multiplets weights no larger than four for the purpose of simplicity. Furthermore, as we have mentioned before, only one lepton-number-violating term, namely, the Majorana mass term of the fermion singlets, will be discussed. Note that other lepton-number-violating mass terms are equally allowed by the modular symmetry itself, and their phenomenological implications can be analyzed in a similar way. The construction starts with the mass term of the fermion singlets, for which we have the following possibilities\footnote{When promoting the fields into chiral superfields, we use the notations $S^{}_\mathrm{L}\rightarrow S$ and $ (N^{}_\mathrm{R})^\mathrm{C}\rightarrow N_{}^\mathrm{C}$.}
\begin{align}
&\textbf{A1}: g \left(S S\right)_{\bf{1}}^{},\quad k_{S}^{}=0\; ;\\
%     (4)
&\textbf{A2}: g \left(S S\right)_{\bf{2}}^{} Y^{}_\mathbf{2},\quad k_{S}^{}= -1\; ;\\
%     (5)
&\textbf{A3}: g \left[ \left(S S\right)_{\bf{1}}^{} Y^{(4)}_\mathbf{1} + r_{g_1}^{} e^{\mathrm{i}p_{g_1}^{}}\left( S S \right)_{\bf{2}}^{} Y^{(4)}_\mathbf{2} + r_{g_2}^{} e^{\mathrm{i}p_{g_2}^{}}\left( S S \right)_{\bf{3}}^{} Y^{(4)}_\mathbf{3} \right],\quad k_{S}^{}=-2\; ,
%     (6)
\end{align}
where $Y_{\bf{2}}^{}$ is a modular-form multiplet of weight $2$, while $Y_{\bf{1}}^{(4)} $, $Y_{\bf{2}}^{(4)} $ and $Y_{\bf{3}}^{(4)}$ are those of weight $4$; $g_{}^{}$ is a scale factor with mass dimension, and $r_{g_i}^{}$ and $p_{g_i}^{}$ (for $i=1,2$) are the relative magnitudes and phases of the relevant terms. Note that $\textbf{Ai}$ (for $i=1,2,3$) refer to different cases for the weight $k_S^{}$, and we list all allowed couplings with a given weight. For a general introduction to modular symmetries, see Appendix~\ref{apd:modular}. More details on these modular-form multiplets can be found in Appendix~\ref{apd:S4}. It is worth mentioning that the Majorana mass term $\left(S S\right)_{\bf{3}^\prime}^{} Y^{}_{\bf{3}^\prime}$ or $\left(S S\right)_{\bf{3}^\prime}^{}Y^{(4)}_{\bf{3}^\prime}$ is not allowed, as the irreducible representation $\bf{3^\prime}$ is anti-symmetric.
Then the coupling between fermion singlets and right-handed neutrinos, together with the Dirac neutrino coupling, read
\begin{align}
&\textbf{B1}: \Lambda \left( S N^\mathrm{C}_{} \right)^{}_{\bf{3}^\prime}  Y^{}_{\mathbf{3}^\prime},\quad k^{}_{N^\mathrm{C}} = -2 - k^{}_{S}\; ;\\
%     (7)
&\textbf{B2}: \Lambda \left[ \left( S N^\mathrm{C}_{} \right)^{}_{\bf{3}}  Y^{(4)}_\mathbf{3} + r_\Lambda^{} e^{\mathrm{i}p_\Lambda^{}}\left( S N^\mathrm{C}_{}\right)^{}_{\bf{3}^\prime} Y^{(4)_{}}_{\mathbf{3}^\prime} \right],\quad k_{N^\mathrm{C}} = -4 - k_{S}^{}\; ;\\
%     (8)
&\textbf{C1}: y \left( L N^\mathrm{C}_{}\right)^{}_{\bf{3}^\prime} Y^{}_{\mathbf{3}^\prime},\quad k_{L}^{} = -2 - k_{N^\mathrm{C}}^{}\; ;\\
%    (9)
&\textbf{C2}: y \left[ \left( L N^\mathrm{C}\right)^{}_{\bf{3}} Y^{(4)}_\mathbf{3} + r_y^{} e^{\mathrm{i}p_y^{}} \left( L N^\mathrm{C}_{} \right)^{}_{\bf{3}^\prime} Y^{(4)}_{\mathbf{3}^\prime} \right],\quad k_{L}^{} = -4 - k_{N^\mathrm{C}}^{} \; ,
%    (10)
\end{align}
where $\textbf{Bi}$ (for $i=1,2$)  denote different choices of the Yukawa coupling between the fermion singlets and right-handed neutrinos, and $\textbf{Ci}$ (for $i=1,2$)  stand for different Dirac neutrino couplings; $\Lambda$ is a scale factor with mass dimension, $y$ is a dimensionless coupling, and $\{r_\Lambda^{}, r_y^{}\}$ and $\{p_\Lambda^{}, ~p_y^{}\}$ are the relative magnitudes and phases of the two relevant terms.

In the charged lepton sector, we take the right-handed charged leptons as singlets of $S_4^{}$. To avoid any degeneracy of charged-lepton masses, we have to introduce three different modular-form triplets to make up the charged-lepton Yukawa term. The superpotential relevant for the charged lepton Yukawa interaction can be written as
\begin{align}
W_l^{} = \alpha \left( L E^\mathrm{C}_1 \right)^{}_{\bf{3}^\prime} Y^{}_{\bf{3}^\prime} H^{}_{\rm d}+ \beta \left( L E^\mathrm{C}_2 \right)^{}_{\bf{3}} Y_{\bf{3}}^{(4)} H^{}_{\rm d}+ \gamma \left( L E^\mathrm{C}_3 \right)^{}_{\bf{3}^\prime} Y_{\bf{3}^\prime}^{(4)} H^{}_{\rm d}\;, \label{eq:Wl}
%    (11)
\end{align}
where dimensionless couplings $\alpha$, $\beta$, $\gamma$ can be chosen to be real without loss of generality.
Starting from the modular forms of the lowest weight, we find $Y^{}_{\bf{3}^\prime}$, $Y_{\bf{3}}^{(4)}$ and $Y_{\bf{3}^\prime}^{(4)}$ as in Eq.~(\ref{eq:Wl}). However, their relative places in the superpotential can be switched, which give rise to the same mass matrix and thus do not affect the final results~\cite{Novichkov:2018ovf}.

%%%%%%%%%%%%%%%%%%%%%%%%%%%%%%%%%%% Table 1 %%%%%%%%%%%%%%%%%%%%%%%%%%%
\begin{table}[t!]
 \caption{\label{tab:assign} The charge assignments of chiral superfields under the SM ${\rm SU}(2)$ gauge symmetry and the modular $S^{}_4$ symmetry.}
 \vspace{0.3cm}
 \centering
  \begin{tabular}{c|c|c|c|c|c|c|c|c}
  \toprule
  \hline
  \xrowht{10pt} & $L$ & $H^{}_{\rm u}$ & $H^{}_{\rm d}$ & $E^{\rm C}_1$ & $E^{\rm C}_2$ & $E^{\rm C}_3$ & $N^\mathrm{C}_{}$ & $S$\\\hline
   \xrowht{10pt} ${\rm SU}(2)$ & 2 & 2       & 2      & 1 & 1 & 1 & 1 & 1 \\
   \xrowht{10pt} $S^{}_4$ & {\bf 3} & {\bf 1}       & {\bf 1}      & ${\bf 1}^\prime$ & {\bf 1} & ${\bf 1}^\prime$ & {\bf 2} & {\bf 3} \\
  \bottomrule
\end{tabular}
\end{table}
%%%%%%%%%%%%%%%%%%%%%%%%%%%%%%%%%%%%%%%%%%%%%%%%%%%%%%%%%%%%%%%%%%%%%%

Since only the neutrino sector has multiple possibilities, we specify a model by choosing one possible coupling in the neutrino sector and labeling it as $\textbf{AiBjCk}$ (for $i=1,2,3$ and $j,k=1,2$). We have twelve models considering different combinations.
The full assignments of the chiral superfields under the SM ${\rm SU}(2)$ gauge symmetry and the modular $S^{}_4$ symmetry are shown in Table~\ref{tab:assign}. Note that once $k_{S}^{}$ is chosen, modular weights of the other fields are fixed accordingly in a model. In addition, all the five distinct irreducible representations of $S^{}_4$ (i.e., $\bf{1}$, $\bf{1^\prime}$, $\bf{2}$, $\bf{3}$ and $\bf{3^\prime}$) are utilized in the model in a natural way.
Given the charge assignments, one can easily write down the gauge- and modular-invariant superpotential for neutrino masses. For example, in model $\textbf{A2B2C1}$, we have
%\vspace{3cm}
\begin{align}
W_\nu^{} &= y \left( L N^\mathrm{C} \right)^{}_{\bf{3}^\prime} Y^{}_{\bf{3}^\prime}  H^{}_{\rm u}+  \Lambda \left[ \left( S N^\mathrm{C}_{}\right)^{}_{\bf{3}} Y_{\bf{3}}^{(4)} + r_\Lambda^{} e^{{\rm i} p^{}_\Lambda} \left( S N^\mathrm{C}_{} \right)^{}_{\bf{3}^\prime} Y_{\bf{3}^\prime}^{(4)} \right] + g \left( SS \right)_{\bf{2}}^{} Y^{}_{\bf{2}}\; . \label{eq:Wnu}
%     (12)
\end{align}
One can choose the parameter $g$ to be real, while in general $y$ and $\Lambda$ are complex.

After breaking of the electroweak gauge symmetry and the flavor symmetry, we get the following mass matrix for the charged leptons
\begin{align}
M^{}_l &= v^{}_{\rm d} \left(  \begin{array} {ccc}
 \alpha Y^{}_3 & -2\beta Y^{}_2 Y^{}_3 & 2\gamma  Y^{}_1 Y^{}_3 \\
 \alpha Y^{}_5 & \beta(\sqrt{3} Y^{}_1 Y^{}_4 + Y^{}_2 Y^{}_5) & \gamma (\sqrt{3} Y^{}_2 Y^{}_4 - Y^{}_1 Y^{}_5) \\
 \alpha Y^{}_4 & \beta(\sqrt{3} Y^{}_1 Y^{}_5 + Y^{}_2 Y^{}_4) & \gamma (\sqrt{3} Y^{}_2 Y^{}_5 - Y^{}_1 Y^{}_4) \\
 \end{array}\right)^* \; ,
 \label{eq:Ml}
%     (13)
\end{align}
where $v^{}_{\rm d} = \langle H^{}_{\rm d} \rangle$ is the vev of the down-type Higgs doublet.
In the neutrino sector, the singlet fermion mass term has the following structures
\begin{align}
\textbf{A1}:\mu  =& ~g^*_{} \left( \begin{array}{ccc}
1  & 0 & 0 \\
0 & 0 & 1 \\
0 & 1 & 0 \\
\end{array} \right)^*\;,\\
%     (14)
\textbf{A2}:\mu  = & ~g^*_{} \left( \begin{array}{ccc}
Y^{}_1  & 0 & 0 \\
0 &  \displaystyle \frac{\sqrt{3}}{2}Y^{}_2 & \displaystyle -\frac{1}{2}Y^{}_1 \\
0 &  \displaystyle -\frac{1}{2}Y^{}_1 & \displaystyle \frac{\sqrt{3}}{2}Y^{}_2 \\
\end{array} \right)^*\;,\\
%     (15)
\textbf{A3}:\mu  =& ~+g^*_{} \left[ \left( \begin{array}{ccc}
Y_1^2 + Y_2^2  & 0 & 0 \\
0 &  0 & Y_1^2 + Y_2^2 \\
0 &  Y_1^2 + Y_2^2 & 0\\
\end{array} \right)^* \right.\nonumber\\
&~+ \left. r_{g_1}^{} e^{-\mathrm{i} p_{g_1}^{}}\left( \begin{array}{ccc}
Y_2^2 - Y_1^2  & 0 & 0 \\
0 &  \displaystyle \sqrt{3} Y^{}_1Y^{}_2 & \displaystyle -\frac{1}{2} \left( Y_2^2 - Y_1^2 \right) \\
0 &  \displaystyle -\frac{1}{2} \left( Y_2^2 - Y_1^2 \right) & \displaystyle \sqrt{3} Y^{}_1Y^{}_2\\
\end{array} \right)^*  \right.\nonumber\\
&~+ \left. r_{g_2}^{} e^{-\mathrm{i} p_{g_2}^{}}\left( \begin{array}{ccc}
0  & \displaystyle -\left( \sqrt{3} Y^{}_1Y^{}_5 + Y^{}_2 Y^{}_4 \right) & \displaystyle \sqrt{3} Y^{}_1Y^{}_4 + Y^{}_2 Y^{}_4  \\
\displaystyle -\left( \sqrt{3} Y^{}_1Y^{}_5 + Y^{}_2 Y^{}_4 \right) &  2Y^{}_2Y^{}_3 & 0 \\
\displaystyle \sqrt{3} Y^{}_1Y^{}_4 + Y^{}_2 Y^{}_4 &  0  & -2Y^{}_2Y^{}_3\\
\end{array} \right)^* \right]\;.
\label{eq:mu}
%     (16)
\end{align}
The singlet-right-handed neutrino coupling reads
\begin{align}
\textbf{B1}:M^{}_{\rm S}  =& ~\Lambda^*_{} \left( \begin{array}{cc}
0    &  - Y^{}_3 \\
\displaystyle \frac{\sqrt{3}}{2} Y^{}_4 & \displaystyle \frac{1}{2} Y^{}_5 \\
\displaystyle \frac{\sqrt{3}}{2} Y^{}_5 & \displaystyle \frac{1}{2} Y^{}_4 \\
\end{array}\right)^* \; ,\\
%     (17)
\textbf{B2}:M^{}_\mathrm{S} = & ~+\Lambda^*_{} \left[\left( \begin{array}{cc}
-2 Y^{}_2 Y^{}_3    & 0 \\
\displaystyle - \frac{1}{2} \left(\sqrt{3} Y^{}_1 Y^{}_4 + Y^{}_2 Y^{}_5 \right) & \displaystyle \frac{\sqrt{3}}{2} \left(\sqrt{3} Y^{}_1 Y^{}_5 + Y^{}_2 Y^{}_4 \right) \\
\displaystyle -\frac{1}{2} \left( \sqrt{3} Y^{}_1 Y^{}_5 + Y^{}_2 Y^{}_4 \right) & \displaystyle \frac{\sqrt{3}}{2} \left( \sqrt{3} Y^{}_1 Y^{}_4 + Y^{}_2 Y^{}_5 \right) \\
\end{array}\right)^* \right. \nonumber \\
& ~ +\left. r_\Lambda^{} e^{-{\rm i} p^{}_\Lambda} \left( \begin{array}{cc}
0    &  -2 Y^{}_1 Y^{}_3\\
\displaystyle \frac{\sqrt{3}}{2} \left( \sqrt{3} Y^{}_2 Y^{}_5 - Y^{}_1 Y^{}_4 \right) & \displaystyle \frac{1}{2} \left( \sqrt{3} Y^{}_2 Y^{}_4 - Y^{}_1 Y^{}_5 \right) \\
\displaystyle \frac{\sqrt{3}}{2} \left( \sqrt{3} Y^{}_2 Y^{}_4 - Y^{}_1 Y^{}_5 \right)  & \displaystyle \frac{1}{2} \left( \sqrt{3} Y^{}_2 Y^{}_5 - Y^{}_1 Y^{}_4 \right) \\
\end{array}\right)^* \right]\; .
\label{eq:Ms}
%     (18)
\end{align}
The Dirac neutrino mass term is
\begin{align}
\textbf{C1}:M^{}_{\rm D}  =& ~y^*_{} v^{}_{\rm u} \left( \begin{array}{cc}
0    &  - Y^{}_3 \\
\displaystyle \frac{\sqrt{3}}{2} Y^{}_4 & \displaystyle \frac{1}{2} Y^{}_5 \\
\displaystyle \frac{\sqrt{3}}{2} Y^{}_5 & \displaystyle \frac{1}{2} Y^{}_4 \\
\end{array}\right)^* \; ,\\
%     (19)
\textbf{C2}:M^{}_{\rm D} = & ~+ y_{}^* v^{}_{\rm u} \left[ \left( \begin{array}{cc}
-2 Y^{}_2 Y^{}_3    & 0 \\
\displaystyle - \frac{1}{2} \left(\sqrt{3} Y^{}_1 Y^{}_4 + Y^{}_2 Y^{}_5 \right) & \displaystyle \frac{\sqrt{3}}{2} \left(\sqrt{3} Y^{}_1 Y^{}_5 + Y^{}_2 Y^{}_4 \right) \\
\displaystyle -\frac{1}{2} \left( \sqrt{3} Y^{}_1 Y^{}_5 + Y^{}_2 Y^{}_4 \right) & \displaystyle \frac{\sqrt{3}}{2} \left( \sqrt{3} Y^{}_1 Y^{}_4 + Y^{}_2 Y^{}_5 \right) \\
\end{array}\right)^* \right. \nonumber \\
& ~+ \left. r_y^{} v^{}_{\rm u} e^{-{\rm i} p^{}_y} \left( \begin{array}{cc}
0    &  -2 Y^{}_1 Y^{}_3\\
\displaystyle \frac{\sqrt{3}}{2} \left( \sqrt{3} Y^{}_2 Y^{}_5 - Y^{}_1 Y^{}_4 \right) & \displaystyle \frac{1}{2} \left( \sqrt{3} Y^{}_2 Y^{}_4 - Y^{}_1 Y^{}_5 \right) \\
\displaystyle \frac{\sqrt{3}}{2} \left( \sqrt{3} Y^{}_2 Y^{}_4 - Y^{}_1 Y^{}_5 \right)  & \displaystyle \frac{1}{2} \left( \sqrt{3} Y^{}_2 Y^{}_5 - Y^{}_1 Y^{}_4 \right) \\
\end{array}\right)^* \right]\; ,
%     (20)
\end{align}
where $v^{}_{\rm u} \equiv \langle H^{}_{\rm u}\rangle$ is the vev of the up-type Higgs doublet.

The $8\times 8$ mass matrix written in the basis $(\nu^{}_\mathrm{L}, N_\mathrm{R}^\mathrm{C}, S_\mathrm{L}^{})$ is
\begin{align}
\mathcal{M} = \left( \begin{array}{ccc}
\textbf{0}^{}_{3\times 3} & \left[ M^{}_\mathrm{D} \right]^{}_{3\times 2}        & \textbf{0}^{}_{3\times 3} \\
\left[M_\mathrm{D}^\mathrm{T}\right]^{}_{2\times 3}   & \textbf{0}_{2\times 2}  & \left[M_\mathrm{S}^\mathrm{T} \right]^{}_{2\times 3}\\
\textbf{0}^{}_{3\times 3} &  \left[ M^{}_\mathrm{S} \right]^{}_{3\times 2}      & \left[ \mu \right]^{}_{3\times 3}\\
\end{array} \right)\;.
\label{eq:Mfull}
%     (21)
\end{align}
After block diagonalization, one gets the light neutrino mass matrix as
\begin{align}
M_\nu^{} &= -M_\mathrm{D}^{}    \left( M_\mathrm{S}^\mathrm{T} M_\mathrm{S}^{} \right)^{-1}  M_\mathrm{S}^\mathrm{T} \mu M_\mathrm{S}^{}    \left( M_\mathrm{S}^\mathrm{T} M_\mathrm{S}^{}\right)^{-1}  M_\mathrm{D}^\mathrm{T}\;.
\label{eq:mnu}
%     (22)
\end{align}
The $M_\nu^{}$ expression looks slightly different from that in the canonical ISS case (see Appendix~\ref{apd:block}), mainly due to the complexity caused by the rectangular shape of $M_\mathrm{D}^{}$ and $M_\mathrm{S}^{}$ matrices. By the canonical ISS model we mean models with $n_{N_\mathrm{R}}^{} = n_{S_\mathrm{L}}^{}$, where $n_{N_\mathrm{R}}^{}$ and $n_{S_\mathrm{L}}^{}$ stand for the number of $N_\mathrm{R}^{}$ and that of $S_\mathrm{L}^{}$ respectively. Details of the block diagonalization procedure can be found in Appendix~\ref{apd:block}.

The light neutrino mass matrix can be diagonalized by a unitary matrix $U$ as $U^\dagger_{} M_\nu^{} U^*_{} =\widehat{M}_\nu^{}$, where $\widehat{M}^{}_\nu \equiv \mathrm{Diag}\{m_1^{},m_2^{},m_3^{}\}$ with $m_i^{}$ being neutrino masses. The neutrino flavor mixing matrix is non-unitary (but we still use the symbol $U_\nu^{}$) and connected to the unitary one $U$ as
\begin{align}
U_\nu^{}= (1-\eta) U\;,
\label{eq:Unu}
%     (23)
\end{align}
where the $\eta$ matrix measures the unitarity violation and is approximately given by $\eta\simeq R R_{}^\dagger/2$. The $R$ matrix comes from the block diagonalization and can be expressed as
\begin{align}
R=\left[
\begin{array}{cc}
M_\mathrm{D}^* \left( M_\mathrm{S}^\dagger M_\mathrm{S}^*M_\mathrm{S}^\dagger M_\mathrm{S}^*\right)^{-1}   M_\mathrm{S}^\dagger \mu^*_{} M_\mathrm{S}^* & - M_\mathrm{D}^*  \left( M_\mathrm{S}^\dagger M_\mathrm{S}^*\right)^{-1} M_\mathrm{S}^\dagger\\
\end{array}
\right]_{ 3\times 5}^{}\;,
%     (24)
\end{align}
which leads to $\eta \simeq M_\mathrm{D}^* \left( M_\mathrm{S}^\dagger M_\mathrm{S}^*\right)^{-1}_{} M_\mathrm{D}^\mathrm{T}/2$, where the higher-order term $\mathcal{O}(\mu^2_{} M_\mathrm{D}^2/M_\mathrm{S}^4)$ has been omitted.
From Eq.~(\ref{eq:mnu}) we see that the light neutrino mass is at $\mathcal{O}(\mu M_\mathrm{D}^2/M_\mathrm{S}^2)$. Considering a hierarchy $\mu \ll M_\mathrm{D}^{} < M_\mathrm{S}^{}$, e.g., $\mu \sim 1$ keV, $M_\mathrm{D}^{} \sim 10^2$ GeV, $M_\mathrm{S}^{} \sim 10^4$ GeV, one can have sub-eV masses for light neutrinos.

\section{Lepton Flavor Mixing}\label{sec:ph}

As neutrino oscillation parameters are precisely measured, we first consider the constraints on the models from neutrino oscillation.
To make a comparison with the observables, we take the leading-order approximation, i.e., taking the effective Majorana neutrino mass matrix in Eq.~(\ref{eq:mnu}) and the charged lepton mass matrix in Eq.~(\ref{eq:Ml}) as the starting point. By diagonalizing both, we get observables as functions of model parameters. As pointed out in the last section, in ISS models, the light neutrino mixing matrix is not unitary. However, the non-unitarity is stringently constrained by current experiments. Ref.~\cite{Ellis:2020hus} offers a global fit of constraints on leptonic unitarity, from which one can see that, even in the agnostic case,\footnote{The agnostic case refers to the matrix $\mathcal{U}$ defined in $\nu_\alpha = \sum_k \mathcal{U}_{\alpha k} \nu_k$ is not unitary. Here $\alpha \in e,~\mu,~\tau,~...$ is the flavor index, $k \in 1,~2,~3,~...$ is the index for mass eigenstates. The agnostic case has the largest non-unitarity effects. ISS models correspond to a submatrix case where the matrix $\mathcal{U}$ is unitary. The non-unitarity in the submatrix case is a subset of that in the agnostic case, as shown in Fig.~13 of Ref.~\cite{Ellis:2020hus}.} the row and column normalizations agree well with one, i.e., the unitary value. The $3\sigma$ credible regions (CR) allow only percent-level deviations from one in the first and second row and a few tens percent of the rest. It is reasonable to assume the unitarity at the leading order.

We perform a Bayesian model comparison among the twenty-four candidate scenarios (twelve models with both mass orderings of the light neutrinos). We construct a seven-dimensional Gaussian likelihood function using observables from Table~\ref{tab:obs}. The number of model parameters is counted as follows: $2$ real parameters from the modulus $\tau$, plus $3$ parameters in the charged lepton sector, plus $N$ parameters in the neutrino sector, where $N$ is model-dependent and is listed in the following:
\begin{align}
&\textbf{A1, A2}: ~0 \nonumber \\
&\textbf{A3}: ~4~(r_{g_1}^{},~r_{g_2}^{},~p_{g_1}^{},~p_{g_2}^{}) \nonumber \\
&\textbf{B1, C1}: ~0 \nonumber \\
&\textbf{B2, C2}: ~2~(r_\Lambda^{}, ~p_\Lambda^{} ~\text{or} ~r_y^{}, ~p_y^{}) \nonumber
\end{align}
With noncommittal model priors, the model selection is made by comparing the Bayesian factors.  We use MultiNest~\cite{Feroz:2007kg,Feroz:2008xx,Feroz:2013hea} to sample and evaluate the Bayesian evidence. The priors are set on the same rooting to avoid any bias on evidence evaluation.
%We set $\mathrm{nlive}=1000,~\mathrm{efr}=0.3, \text{and}~\mathrm{tol}=0.5$.
We find that among the $24$ candidate scenarios, there is strong preference for normal ordering of \textbf{A1B2C2}. In the following, we explore the parameter space of model \textbf{A1B2C2} in detail to see if it can be in accordance with the oscillation data.

%%%%%%%%%%%%%%%%%%%%%%%%%%%%%%%%%%% Table 2 %%%%%%%%%%%%%%%%%%%%%%%%%%%
\begin{table}[t!]
 \caption{\label{tab:obs} The observables in the charged lepton and neutrino sector. The charged lepton Yukawa couplings are evaluated at the $m_Z$ scale, taken from Ref.~\cite{Huang:2020hdv}. The neutrino oscillation parameters are taken from NuFIT 5.0 (2020)~\cite{Esteban:2020cvm}. The average errors are shown in the brackets. }
 \vspace{0.3cm}
 \centering
  \begin{tabular}{c|c|c}
  \toprule
  \hline
  \xrowht{10pt} $y^{}_e$     & \multicolumn{2}{c}{$2.7771(26)\times 10^{-6}$}\\
  \xrowht{10pt} $y^{}_\mu$ & \multicolumn{2}{c}{$5.8504(13)\times 10^{-4}$}\\
  \xrowht{10pt} $y^{}_\tau$ & \multicolumn{2}{c}{$9.9372(16)\times 10^{-3}$}\\
  \hline
   \xrowht{10pt} & NO & IO\\
   \hline
  \xrowht{10pt} $\sin^2\theta_{12}$ & $0.304(13)$      & $0.304(13)$ \\
  \xrowht{10pt} $\sin^2\theta_{13}$ & $0.02221(65)$  & $0.02240(62)$\\
  \xrowht{10pt} $\sin^2\theta_{23}$ & $0.570(21)$      &  $0.575(19)$\\
  \xrowht{10pt} $\displaystyle r \equiv \frac{\Delta m^2_{21}}{|\Delta m^2_{31(2)}|}$ & $0.0295(9)$ & $0.0297(9)$\\
  \bottomrule
\end{tabular}
\end{table}
%%%%%%%%%%%%%%%%%%%%%%%%%%%%%%%%%%%%%%%%%%%%%%%%%%%%%%%%%%%%%%%%%%%%

%%%%%%%%%%%%%%%%%%%%%%%%%%%%%%%%% Table 3 %%%%%%%%%%%%%%%%%%%%%%%%%%%%%%%%%%%
\begin{table}[t!]
 \caption{\label{tab:parambft} The best-fit values of the model parameters.}
 \vspace{0.3cm}
 \centering
  \begin{tabular}{c|c|c|c|c|c|c|c|c}
  \toprule
  \hline
  \xrowht{10pt} $\mathrm{Re}\tau$ & $\mathrm{Im}\tau$ & $\alpha$ & $\beta$ & $\gamma$ & $r_\Lambda^{}$ & $r_y^{}$ & $p_\Lambda^{}$ & $p_y^{}$\\
  \hline
  \xrowht{10pt} $0.260$ & $1.762$ & $7.110\times 10^{-4}$ & $1.719\times 10^{-2}$ & $2.999\times 10^{-6}$ & $0.847$ & $0.312$ & $3.913$ & $2.370$\\
  \bottomrule
\end{tabular}
\end{table}
%%%%%%%%%%%%%%%%%%%%%%%%%%%%%%%%%%%%%%%%%%%%%%%%%%%%%%%%%%%%%%%%%%%%

%%%%%%%%%%%%%%%%%%%%%%%%%%%%%%%%% Table 4 %%%%%%%%%%%%%%%%%%%%%%%%%%%%%%%%%%%
\begin{table}[t!]
 \caption{\label{tab:bft} The observables calculated from the best-fit parameters and the predictions for the phases.}
 \vspace{0.3cm}
 \centering
  \begin{tabular}{c|c}
  \toprule
  \hline\xrowht{10pt}
  Parameters & Predicted value \\
  \hline
  \xrowht{10pt} $\sin^2\theta_{12}$ & $0.305$      \\
  \xrowht{10pt} $\sin^2\theta_{13}$ & $0.02227$ \\
  \xrowht{10pt} $\sin^2\theta_{23}$ & $0.571$       \\
  \xrowht{10pt} $\displaystyle \frac{\Delta m^2_{21}}{\Delta m^2_{31}}$ & $0.0296$ \\
  \xrowht{10pt} $y^{}_e$      & $ 2.7774\times 10^{-6}$  \\
  \xrowht{10pt} $y^{}_\mu$ & $5.8505\times 10^{-4}$  \\
  \xrowht{10pt} $y^{}_\tau$ & $9.9372 \times 10^{-3}$ \\
  \hline
  \xrowht{10pt} $\delta~[ ^\circ]$ & $108.9$\\
  \xrowht{10pt} $\alpha_{21}~[ ^\circ]$ & $25.8$\\
  \xrowht{10pt} $\alpha_{31}~[ ^\circ]$ & $52.7$\\
  \bottomrule
\end{tabular}
\end{table}
%%%%%%%%%%%%%%%%%%%%%%%%%%%%%%%%%%%%%%%%%%%%%%%%%%%%%%%%%%%%%%%%%%%%

%%%%%%%%%%%%%%%%%%%%%%%%%%%%%%% Figure 1 %%%%%%%%%%%%%%%%%%%%%%%%%%%%%%%%%%%%%
\begin{figure}[t!]
\centering
\includegraphics[width=.8\textwidth]{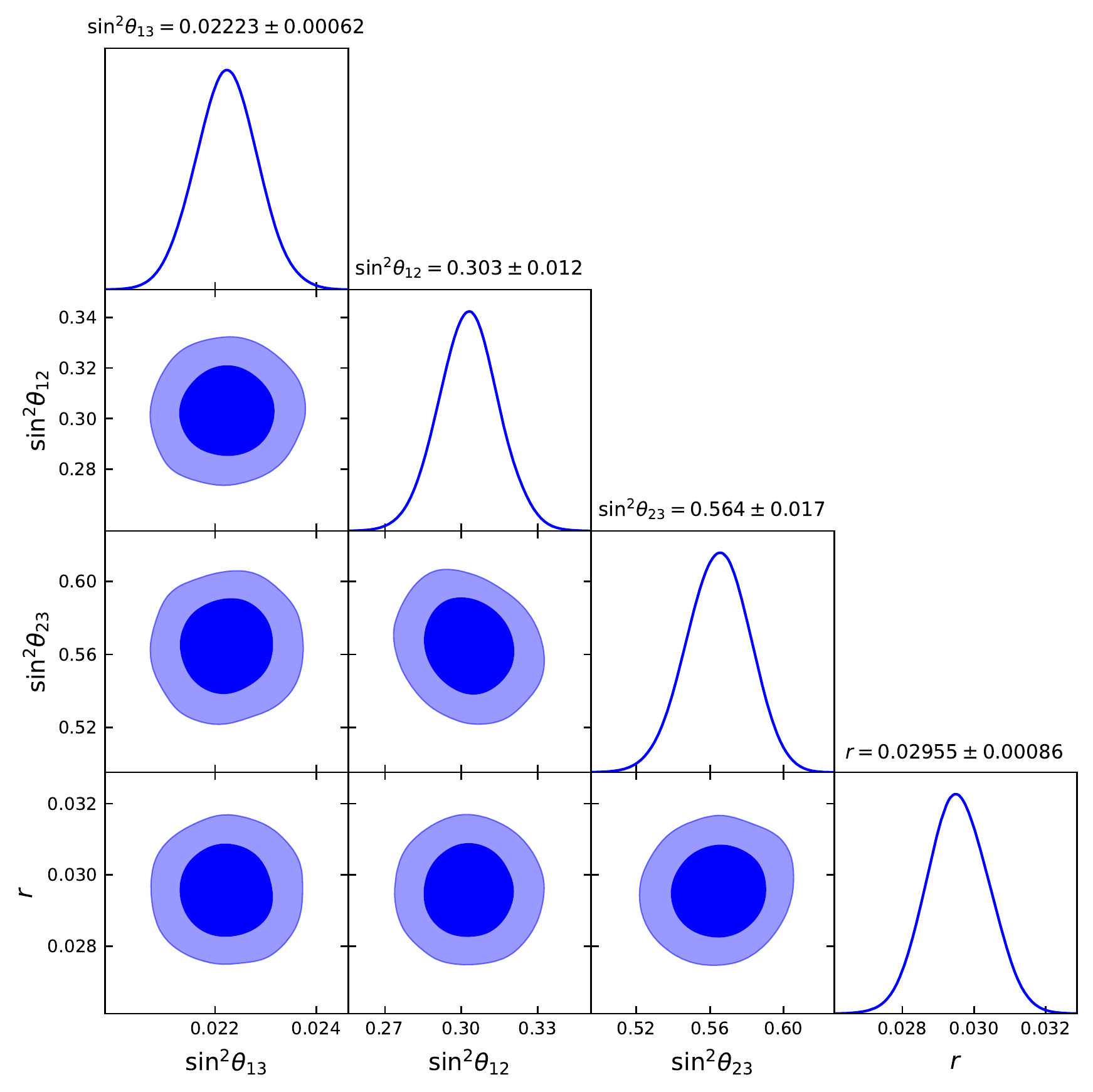}
\caption{ The one- and two-dimensional posterior distribution of the model \textbf{A1B2C2} observables in the neutrino sector. The light and dark blue region correspond to $68\%$ and $95\%$ CR. We also show the mean values with $1\sigma$ error. The plot is generated using GetDist~\cite{Lewis:2019xzd}.}
\label{fig:obs}
%MN root: iss_a1b2c2_p7_
\end{figure}
%%%%%%%%%%%%%%%%%%%%%%%%%%%%%%%%%%%%%%%%%%%%%%%%%%%%%%%%%%%%%%%%%%%%%%%%%%%

%%%%%%%%%%%%%%%%%%%%%%%%%%%%%%% Figure 2 %%%%%%%%%%%%%%%%%%%%%%%%%%%%%%%%%%%%%
\begin{figure}[t!]
\centering
\includegraphics[width=\textwidth]{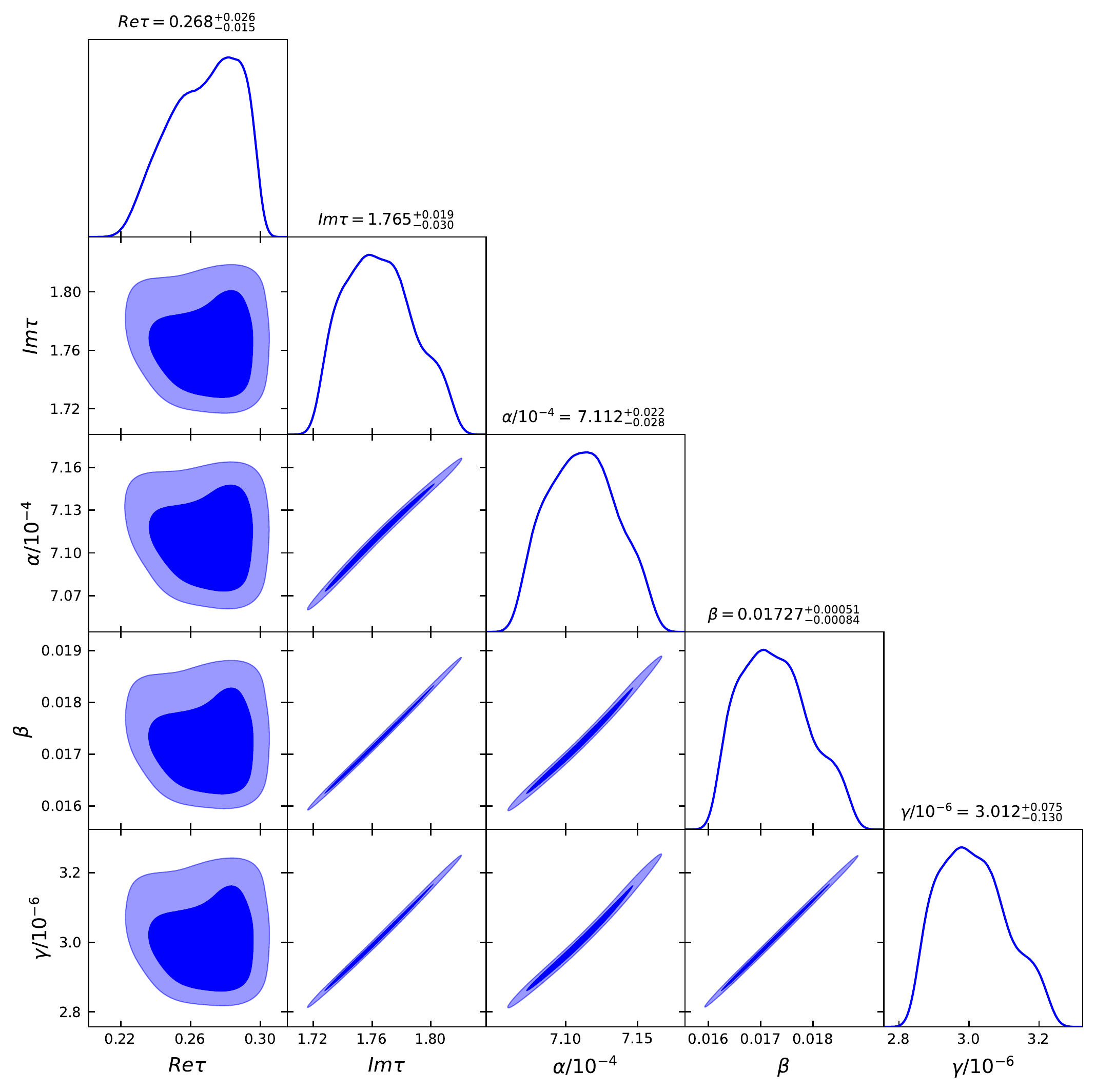}
\caption{ The one- and two-dimensional posterior distribution of the model \textbf{A1B2C2} parameters. The light and dark blue region correspond to $68\%$ and $95\%$ CR. We also show the mean parameter values with $1\sigma$ error. The plot is generated using GetDist~\cite{Lewis:2019xzd}.}
\label{fig:param}
%MN root: iss_a1b2c2_p7_
\end{figure}
%%%%%%%%%%%%%%%%%%%%%%%%%%%%%%%%%%%%%%%%%%%%%%%%%%%%%%%%%%%%%%%%%%%%

Running MultiNest in a parameter-estimation mode, we find the best-fit values of parameters (shown in Table~\ref{tab:parambft}) correspond to $\chi^2_\mathrm{min} \simeq 0.04$, which shows a remarkable agreement with experimental observations. The values of the observables and predictions for the phases corresponding to the best-fit parameters are given in Table~\ref{tab:bft}. We also show the one- and two-dimensional posterior distributions of the model observables in the neutrino sector in Fig.~\ref{fig:obs}. From these results, we see an excellent agreement with the constraining observables. At this stage, the lepton mixing is totally solved. The predicted value for the Dirac CP-violating phase is within $3\sigma$ range. We also have predictions on the two Majorana CP-violating phases defined as follows
\begin{align}
U=
\left(
\begin{array}{ccc}
c^{}_{12}c^{}_{13} & s^{}_{12}c^{}_{13} & s^{}_{13}e^{-\mathrm{i}\delta}         \\
-s^{}_{12}c^{}_{23}-c^{}_{12}s^{}_{23}s^{}_{13}e^{\mathrm{i}\delta} & c^{}_{12}c^{}_{23}-s^{}_{12}s^{}_{23}s^{}_{13}e^{\mathrm{i}\delta} & s^{}_{23}c^{}_{13} \\
s^{}_{12}s^{}_{23}-c^{}_{12}c^{}_{23}s^{}_{13}e^{\mathrm{i}\delta} & -c^{}_{12}s^{}_{23}-s^{}_{12}c^{}_{23}s^{}_{13}e^{\mathrm{i}\delta} & c^{}_{23}c^{}_{13}\\
\end{array}
\right)
\left(
\begin{matrix}
1 & 0 & 0 \cr
0 & e^{\mathrm{i}\alpha^{}_{21}/2} & 0 \cr
0 & 0 & e^{\mathrm{i}\alpha^{}_{31}/2} \cr
\end{matrix}
\right)\;.
%       (25)
\end{align}
The predicted $3\sigma$ CR are
\begin{align}
\delta \in [69.3,139.8]^\circ,\quad \alpha^{}_{21} \in [-18.9,108.3]^\circ,\quad \alpha^{}_{31} \in [-53.9,123.2]^\circ \;,
%       (26)
\end{align}
and the best-fit values are shown in Table~\ref{tab:bft}.

Although the number of model parameters is larger than that of the observables, it is not a trivial fit as the actual number of degrees of freedom is small. As can be seen from the two-dimensional posterior distribution of the model parameter in Fig.~\ref{fig:param}, the parameters $\alpha,~\beta,~\gamma$ are strongly correlated with $\mathrm{Im}\tau$, reducing the effective number of degrees of freedom to six.

The strong correlation of the parameters can be understood using analytical approximation, which is applicable given that the best-fit value of $\mathrm{Im}\tau $ is large. We introduce
$$x\equiv\mathrm{exp}(-\pi \mathrm{Im} \tau/2),\quad y \equiv \pi \mathrm{Re}\tau/2\;,$$
such that $q^{1/4}_{} =x e^{\mathrm{i}y}$. Given the best-fit value $\mathrm{Im}\tau = 1.762$, we find that $x \simeq 0.063$ can be a good perturbative parameter. Expanding the basis in $x$, at the leading order, we get
\begin{align}
Y^{}_1&\simeq -3\pi/8\; ; \\
Y^{}_2&\simeq 3\sqrt{3}\pi x^2 e^{2\mathrm{i}y}\; ; \\
Y^{}_3&\simeq \pi/4\; ; \\
Y^{}_4&\simeq -\sqrt{2}\pi x e^{\mathrm{i}y}\; ; \\
Y^{}_5&\simeq -4\sqrt{2}\pi x^3 e^{3\mathrm{i}y}\;.
%         (27-31)
\end{align}
The charged lepton mass matrix at the leading order reads
\begin{align}
M^{}_l \simeq v^{}_\mathrm{d}
\left( \begin{array}{ccc}
\displaystyle \frac{1}{4} \pi \alpha  & 0 & -\displaystyle \frac{3}{16} \pi^2\gamma\\
0 &  0 & 0\\
0 &  0 & 0 \\
\end{array} \right) + v^{}_\mathrm{d} x e^{-\mathrm{i}y}
\left( \begin{array}{ccc}
0  & 0 & 0\\
0 &  \displaystyle \frac{3}{4} \sqrt{\frac{3}{2}} \pi^2 \beta  & 0\\
-\sqrt{2} \pi \alpha &  0 & -\displaystyle \frac{3}{4\sqrt{2}} \pi^2 \gamma\\
\end{array} \right) \;.
%         (32)
\end{align}
To find $U^{}_l$, it is convenient to use
\begin{align}
M^{}_l M_l^\dagger \simeq ~&v_\mathrm{d}^2
\left( \begin{array}{ccc}
\displaystyle \frac{\pi^2}{256}  (16\alpha^2 + 9\pi^2 \gamma^2)  & 0 & \displaystyle \frac{\pi^2}{64\sqrt{2}}  (-32\alpha^2 + 9\pi^2\gamma^2) xe^{\mathrm{i}y}\\
0 &  \displaystyle \frac{27}{32} \pi^4 \beta^2 x^2  & 0\\
\displaystyle \frac{\pi^2}{64\sqrt{2}}  (-32\alpha^2 + 9\pi^2\gamma^2) xe^{-\mathrm{i}y}&  0 & \displaystyle \frac{\pi^2}{32}  (64\alpha^2 + 9\pi^2 \gamma^2) x^2 \\
\end{array} \right) \;,
%         (33)
\label{eq:Hl}
\end{align}
where we neglect higher-order terms $\mathcal{O}(x^4)$. One can see that the charged lepton sector contributes a $1$-$3$ rotation up to $\mathcal{O}(x^2)$. To fit the charged lepton Yukawas, $x$ or $\mathrm{Im}\tau$ is strongly correlated with $\alpha,~\beta,~\gamma$, as can be seen from the eigenvalues of $M^{}_l M_l^\dagger$, i.e.,
\begin{align}
y_e^2 &\simeq \displaystyle \frac{81 \pi ^4 \alpha ^2 \gamma ^2 x^2}{32 \alpha ^2+18 \pi ^2 \gamma ^2} \;;\\
%        (34)
y_\mu^2 &\simeq \displaystyle \frac{27}{32} \pi ^4 \beta ^2 x^2 \;; \\
%        (35)
y_\tau^2 &\simeq \displaystyle \frac{1}{256} \pi ^2 \left(16 \alpha ^2+9 \pi ^2 \gamma ^2\right)+\frac{ \left(32 \pi  \alpha ^2-9 \pi ^3 \gamma ^2\right)^2 x^2}{32 \left(16 \alpha ^2+9 \pi ^2 \gamma ^2\right)}\;,
%        (36)
\end{align}
where the higher-order contributions are neglected.
Taking the leading contribution and using the best-fit value of $x$ allow a rough estimation of $\alpha \sim \mathcal{O}(10^{-2}),~ \beta \sim \mathcal{O}(10^{-3}),~\gamma \sim \mathcal{O}(10^{-6})$. These magnitudes roughly agree with those we get from the best fit. The point is that $x$, together with the three coefficients $\alpha,~\beta,~\gamma$, are fixed by the charged lepton mass spectrum, so they are correlated as shown in Fig.~\ref{fig:param}.

To estimate the magnitude of the $1$-$3$ rotation, we introduce $H\equiv M^{}_l M_l^\dagger$, and use the fact that $U_l^\dagger H U_l^{} = \mathrm{Diag} \{ m_e^2,m_\mu^2,m_\tau^2\}$, where $U^{}_l$ is parameterized in the standard way with only a nonzero rotation angle $\theta_{13}^l$. We have
\begin{align}
H^{}_{11} &= \displaystyle \frac{1}{2} \left[m_e^2+m_\tau^2 - (m_\tau^2-m_e^2) \cos 2\theta_{13}^l \right] \;; \\
H^{}_{33} &= \displaystyle \frac{1}{2} \left[m_e^2+m_\tau^2 + (m_\tau^2-m_e^2) \cos 2\theta_{13}^l \right] \;; \\
|H^{}_{13}| &= |H^{}_{31}|  = \displaystyle \frac{1}{2} \left(m_\tau^2-m_e^2 \right) \sin 2\theta_{13}^l \;.
%       (37-39)
\end{align}
Looking back at Eq.~(\ref{eq:Hl}), we find
\begin{align}
\tan 2\theta_{13}^l = \displaystyle \frac{2|H^{}_{13}|}{H^{}_{33}-H^{}_{11}}  \simeq 8\sqrt{2} x \;,
%      (40)
\end{align}
which renders $\theta_{13}^l \simeq 18^\circ$ with the best-fit value of $x$. This value is comparable to the one get from the best fit, i.e., $10^\circ$.

The light neutrino mass matrix reads
\begin{align}
M^{}_\nu \simeq
&-
\left( \begin{array}{ccc}
\displaystyle \frac{r_y^2}{r_\Lambda^2}  & 0 & 0\\
0 &  0 & 0\\
0 &  0 & 0\\
\end{array} \right) \nonumber\\
 &+x e^{-2\mathrm{i}y }\displaystyle \frac{\sqrt{2} r^{}_y}{r_\Lambda^2}
\left( \begin{array}{ccc}
0  & *& *\\
\displaystyle \frac{( r^{}_\Lambda-3e^{\mathrm{i}p^{}_\Lambda} ) (r^{}_y+e^{\mathrm{i}p^{}_y}) }{  r^{}_\Lambda+e^{\mathrm{i}p^{}_\Lambda}  } &  0 & 0\\
 r^{}_y-3e^{\mathrm{i}p^{}_y}   &  0 & 0\\
\end{array} \right) \nonumber\\
&- x^2e^{-2\mathrm{i}y}\displaystyle \frac{2}{r_\Lambda^2}
\left( \begin{array}{ccc}
\displaystyle \frac{2 \left(r^{}_\Lambda-3 e^{\mathrm{i} p^{}_\Lambda}\right)^2 r_y^2 }{r_\Lambda^2} & 0 & 0\\
0&  0 & *\\
0 &   \displaystyle \frac{ \left(r^{}_\Lambda-3 e^{\mathrm{i} p^{}_\Lambda}\right) \left(r^{}_y-3 e^{\mathrm{i} p^{}_\Lambda}\right) \left(r^{}_y+e^{\mathrm{i} p^{}_y}\right) }{ r^{}_\Lambda+e^{\mathrm{i} p^{}_\Lambda} } & \displaystyle  \left(r^{}_y-3 e^{\mathrm{i} p^{}_y}\right)^2 \\
\end{array} \right)  \;,~~~~~
%         (41)
\end{align}
where ``$*$ "denotes the symmetric counterpart, we neglect the overall unphysical phase $e^{2\mathrm{i} (p_\Lambda^{}-p_y^{})}$, and we omit the higher-order terms starting from $\mathcal{O}(x^3)$. With the help of the best-fit values of parameters, we find the non-vanishing entities in $M^{}_\nu$ up to $\mathcal{O}(x^2)$ are all of the same magnitude, which makes it difficult to proceed with the analytical approximation. It takes three relatively large rotations to be transformed into diagonal, which is verified by the best-fit parameter values. We observe that the $(\mu,~\mu)$ entity in $M^{}_\nu$ is vanishing. As it is a function of all the mixing angles and phases, among which two Majorana phases are completely unknown, and the Dirac phase is loosely constrained, it is hard to draw any conclusion from this condition.

Note that by far, the full neutrino mass spectrum is not shown, which is only available when we specify the scales of each sector, namely, the coefficients of the matrices $M^{}_\mathrm{D}, ~M^{}_\mathrm{S}, ~\mu$. With these values, we can solve the full neutrino mass matrix in Eq.~(\ref{eq:Mfull}) and determine the active-sterile neutrino mixing. This analysis is performed in the next section, and we provide the conclusion here: the absolute light neutrino mass scale in this model is $m^{}_\mathrm{min} \in [10^{-4},0.1]$ eV, by constraints from both lepton flavor mixing and the keV sterile neutrino dark matter. A large portion of this range is within the aggressive combined (oscillation $+$ non-oscillation) constraint, and nearly all this range is allowed by the conservative combined constraint~\cite{Capozzi:2017ipn}.

\section{keV Sterile Neutrino}\label{sec:dm}

\subsection{Warm Dark Matter}\label{subsec:wdm}
In this subsection, we briefly review the basics of keV-mass sterile neutrinos as a candidate for warm dark matter (WDM). We start with an introduction to the stability and mass scale, and to possible production mechanisms, and then discuss observational constraints from astrophysical $X$-ray search and Lyman-$\alpha$ forest data.

\subsubsection{Stability and Mass Scale}

The basic requirements for a particle dark matter candidate are electrically neutral and stable. Mixing with light active neutrinos, the sterile state is not absolutely stable. It can decay into a light active neutrino and a photon, or three neutrinos. Its lifetime is estimated from the inverse of the total decay rate, which is found to be larger than the age of the Universe as long as the mixing with active neutrinos is small.

Being a fermion dark matter candidate, in our case a sterile neutrino, its mass gets a lower bound by requiring that the maximal kinetic energy it can have is no larger than the gravitational potential energy, i.e., it is gravitationally bounded. From the observed dwarf satellite galaxy mass and radius, one gets the Tremaine-Gunn bound~\cite{Tremaine:1979we},
\begin{align}
m^{}_s \gtrsim 0.5 ~\mathrm{keV}\;.
%         (42)
\end{align}
This is the first argument that sets the sterile neutrino dark matter mass to be keV.

It is a generic feature of ISS models that when $n^{}_{S_\mathrm{L} }>n^{}_{N_\mathrm{R}}$ there will be ($n^{}_{S_\mathrm{L}}- n^{}_{N_\mathrm{R}}$) intermediate-mass states at the scale $\mu$~\cite{Abada:2014vea}. Actually, the whole mass spectrum for neutrinos including the steriles is
\begin{itemize}
\item $n^{}_{\nu_\mathrm{L}}$ light active neutrinos~$\mathcal{O}(\mu M_\mathrm{D}^2/M_\mathrm{S}^2)$
\item ($n^{}_{S_\mathrm{L}}- n^{}_{N_\mathrm{R}}$) light sterile states~$\mathcal{O}(\mu)$
\item $2n^{}_{N_\mathrm{R}}$ heavy states~$\mathcal{O}(M^{}_\mathrm{S})$, which form $n^{}_{N_\mathrm{R}}$ pseudo-Dirac pairs
%with mass differences~$\mathcal{O}(\mu)$
\end{itemize}
Given $\mu \sim 1$ keV, $M^{}_\mathrm{D} \sim 10^2$ GeV and $M^{}_\mathrm{S} \sim 10^4$ GeV, we expect a sterile neutrino at the keV scale. The detailed numerical analysis of the mass spectrum will be given in next subsection.

\subsubsection{Production Mechanism}
In general, keV sterile neutrinos can be produced in the following ways:
\begin{itemize}
\item Dodelson-Widrow (DW) mechanism~\cite{Dodelson:1993je}, in which the sterile neutrino is produced through active-sterile transition at $T \sim 100 $ MeV in the primordial plasma, which is always allowed as long as the active-sterile mixing exists.
\item  Shi-Fuller (SF) mechanism~\cite{Shi:1998km}, in which a pre-existing lepton number asymmetry produces an enhancement of the active-sterile neutrino transition rate, which is in close analog with the Mikheyev-Smirnov-Wolfenstein matter effect~\cite{Wolfenstein:1977ue,Mikheev:1986gs,Mikheev:1986wj} and is also known as the resonant production mechanism.
\end{itemize}
As the sterile neutrinos can only interact feebly (by definition), they cannot be produced in thermal equilibrium. Both of the aforementioned production mechanisms produce a non-thermal momentum distribution. While the DW mechanism gives rise to the thermal distribution with a suppression factor, the SF mechanism usually leads to a ``colder" distribution.
There exist other mechanisms for keV sterile neutrino production (see Refs.~\cite{Abazajian:2001nj,Kusenko:2009up,Merle:2013gea,Merle:2017dhf,Abazajian:2017tcc} for a general introduction), which require either extra particles or extended gauge symmetries, thus are not applicable to our case.

\subsubsection{Constraints from Astrophysics and Structure Formation}

To be a viable dark matter candidate, keV sterile neutrinos have to survive several observational constraints. The first one is the dark matter abundance observed today, which is $\Omega^{}_{\mathrm{DM}} = 0.120 \pm 0.001$ given by Planck~\cite{Aghanim:2018eyx}. The dark matter abundance imposes limits on both dark matter mass and active-sterile mixing, since the dark matter should not be overproduced to overclose the Universe. It is estimated as~\cite{Asaka:2006nq}
\begin{align}
\Omega^{}_{\mathrm{DM}} h^2 = 1.1\times 10^7 \sum_\alpha C^{}_\alpha(m^{}_s) |U^{}_{\alpha s}|^2 \left( \frac{m^{}_s}{\mathrm{keV}} \right)^2, \quad \alpha=e, \mu, \tau\;, \label{eq:abundance}
%         (43)
\end{align}
where $C^{}_\alpha$ are flavor-dependent coefficients whose exact values can be obtained by solving the Boltzmann equations.
It is also possible that sterile neutrinos only serve as a fraction of the total dark matter. In this case, one can introduce an abundance fraction $f^{}_\mathrm{s}=\Omega^{}_\mathrm{s}/\Omega^{}_{\mathrm{DM}}$, and $\Omega^{}_{\mathrm{DM}}$ in Eq.~(\ref{eq:abundance}) should be replaced by $f^{}_\mathrm{s}\Omega^{}_{\mathrm{DM}}$.

Given the mixing with active neutrinos, the sterile neutrinos undergo radiative decays $N \rightarrow \nu \gamma$, which produces a mono-chromatic photon with energy $E^{}_\gamma \simeq m^{}_s / 2$. Such an $X$-ray line has been searched by Chandra~\cite{Horiuchi:2013noa}, XMM-Newton~\cite{Malyshev:2014xqa} and NuSTAR~\cite{Roach:2019ctw}. The non-observation of the $X$-ray puts stringent limits on sterile neutrino mass and mixing.

Being relatively light, sterile neutrinos feature a relatively large free-streaming horizon, which suppresses the small-scale structure of the matter power spectrum, thus get constrained from structure formation. Lyman-$\alpha$ forest data, coming from observing the Lyman-$\alpha$ transition in hydrogen gas of the Universe, gives the intergalactic medium distribution, from which the WDM velocity dispersion can be extracted. Current Lyman-$\alpha$ bound at the $95\%$ confidence level on the WDM mass is~\cite{Viel:2013fqw,Baur:2015jsy,Irsic:2017ixq,Palanque-Delabrouille:2019iyz,Garzilli:2019qki}
\begin{align}
m^{}_{\mathrm{WDM}} \gtrsim \left( 1.9 - 5.3 \right)~\mathrm{keV}\;. \label{eq:Lya}
%         (44)
\end{align}

There are other constraints, e.g., Milky Way satellite galaxies counting~\cite{Cherry:2017dwu} and supernova bounds~\cite{Dolgov:2000jw,Raffelt:2011nc,Zhou:2015jha,Arguelles:2016uwb,Suliga:2020vpz}. However, these constraints are not competitive with the $X$-ray bounds and the Lyman-$\alpha$ constraints, so we do not include them. It is worth mentioning that the astrophysical $X$-ray constraints are model-independent, while the structure formation constraints in Eq.~(\ref{eq:Lya}) from Lyman-$\alpha$ forest data only directly apply to WDM with a thermal distribution.

\subsection{Numerical Results}

%%%%%%%%%%%%%%%%%%%%%%%%%%%%%%% Figure 3 %%%%%%%%%%%%%%%%%%%%%%%%%%%%%%%%%%%%%
\begin{figure}[t!]
\centering
\includegraphics[width=.7\textwidth]{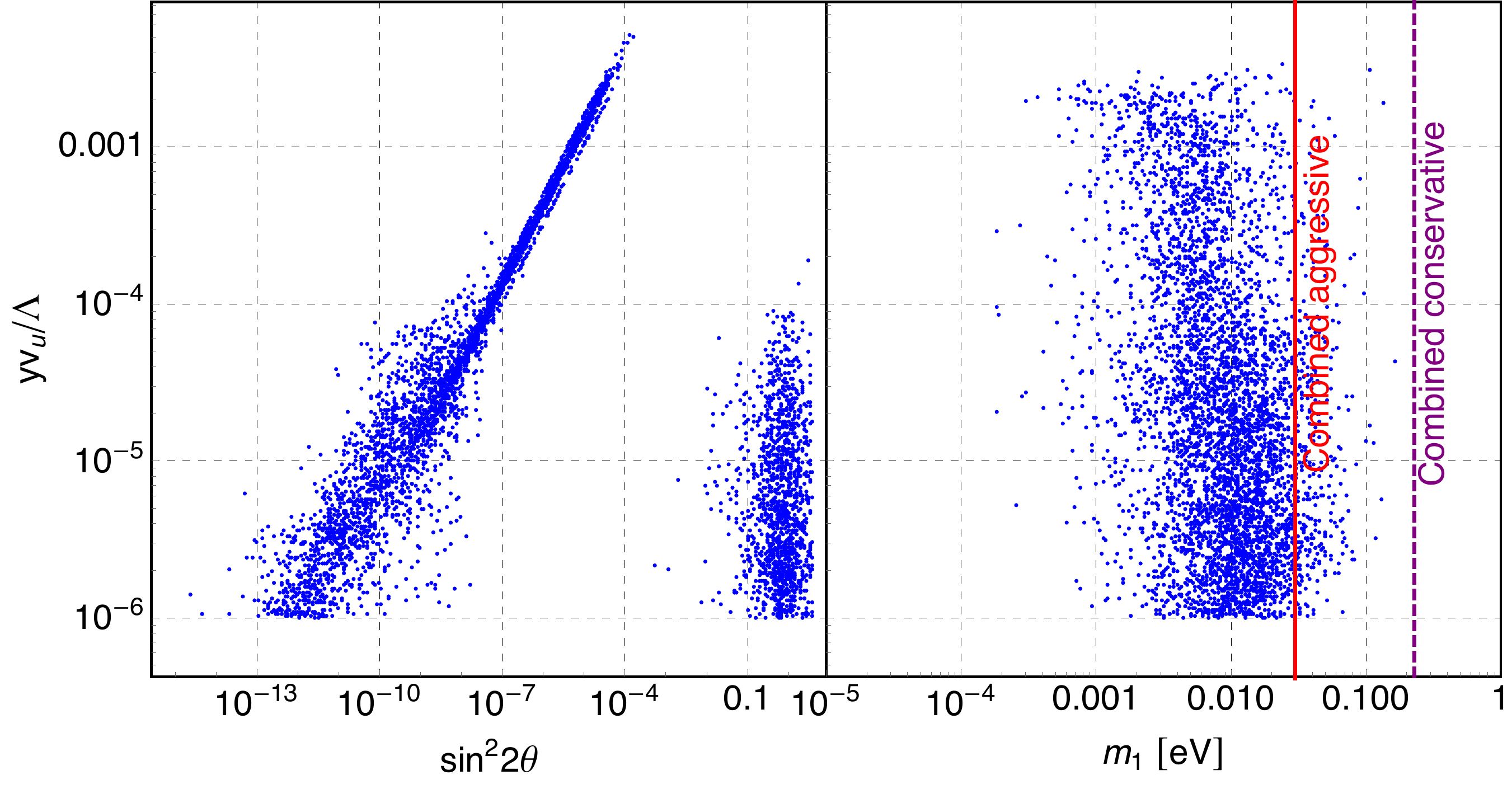}
\includegraphics[width=.7\textwidth]{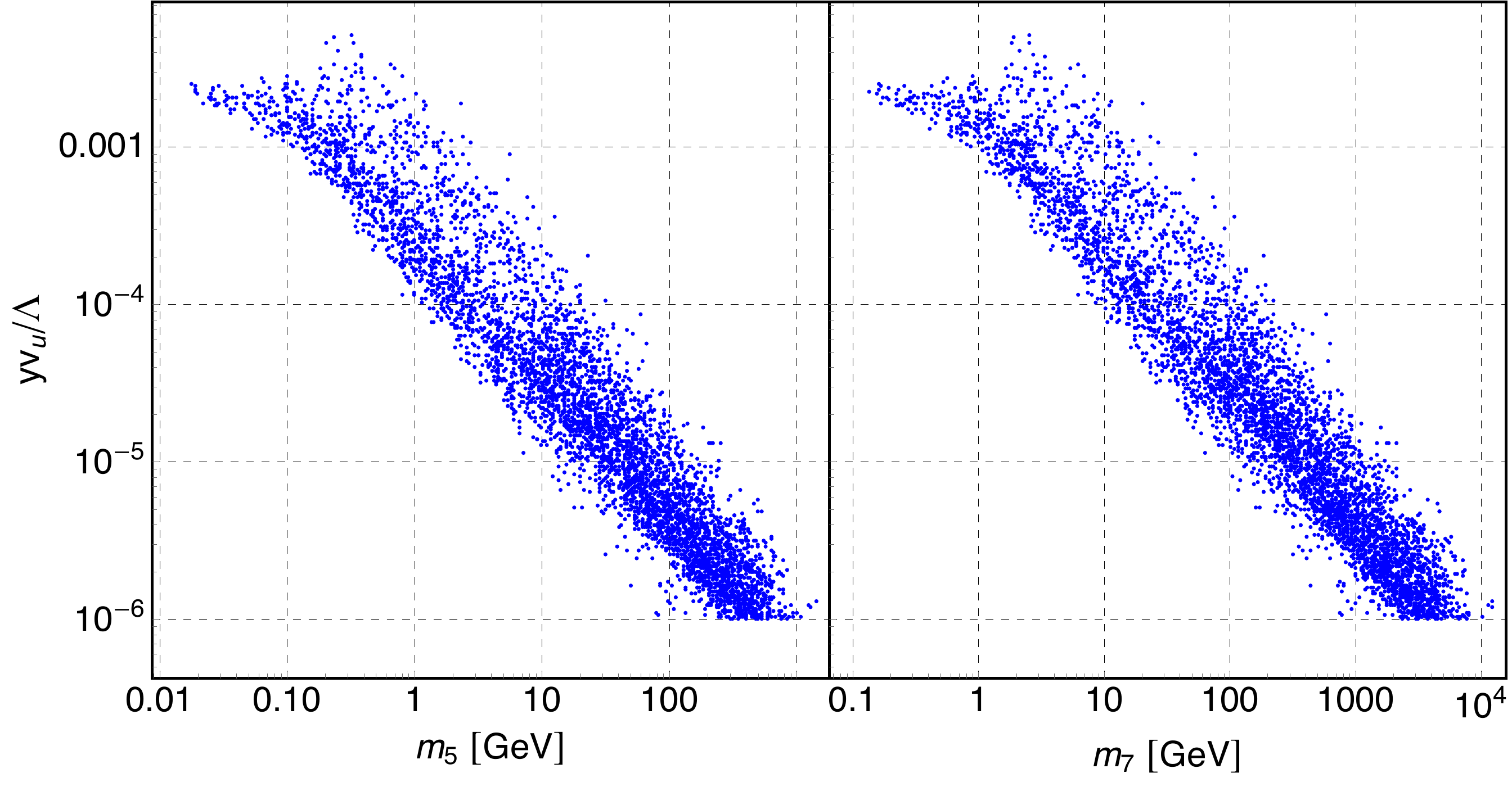}
\caption{The correlation of the scale ratio $yv^{}_\mathrm{u}/\Lambda$ with the active-sterile mixing, the lightest neutrino mass, and the two heavier sterile neutrino masses. The points are selected from the scan with the light sterile mass $m^{}_s \in [0.5,50]$~keV. In the upper-right plot, we show the aggressive (red line) and conservative (purple dashed line) combined (oscillation $+$ non-oscillation) constraints from Ref.~\cite{Capozzi:2017ipn}.}
\label{fig:mms}
\end{figure}
%%%%%%%%%%%%%%%%%%%%%%%%%%%%%%%%%%%%%%%%%%%%%%%%%%%%%%%%%%%%%%%%%%%%%%%%%%

%%%%%%%%%%%%%%%%%%%%%%%%%%%%%%% Figure 4 %%%%%%%%%%%%%%%%%%%%%%%%%%%%%%%%%%%%%
\begin{figure}[t!]
\centering
\includegraphics[width=.5\textwidth]{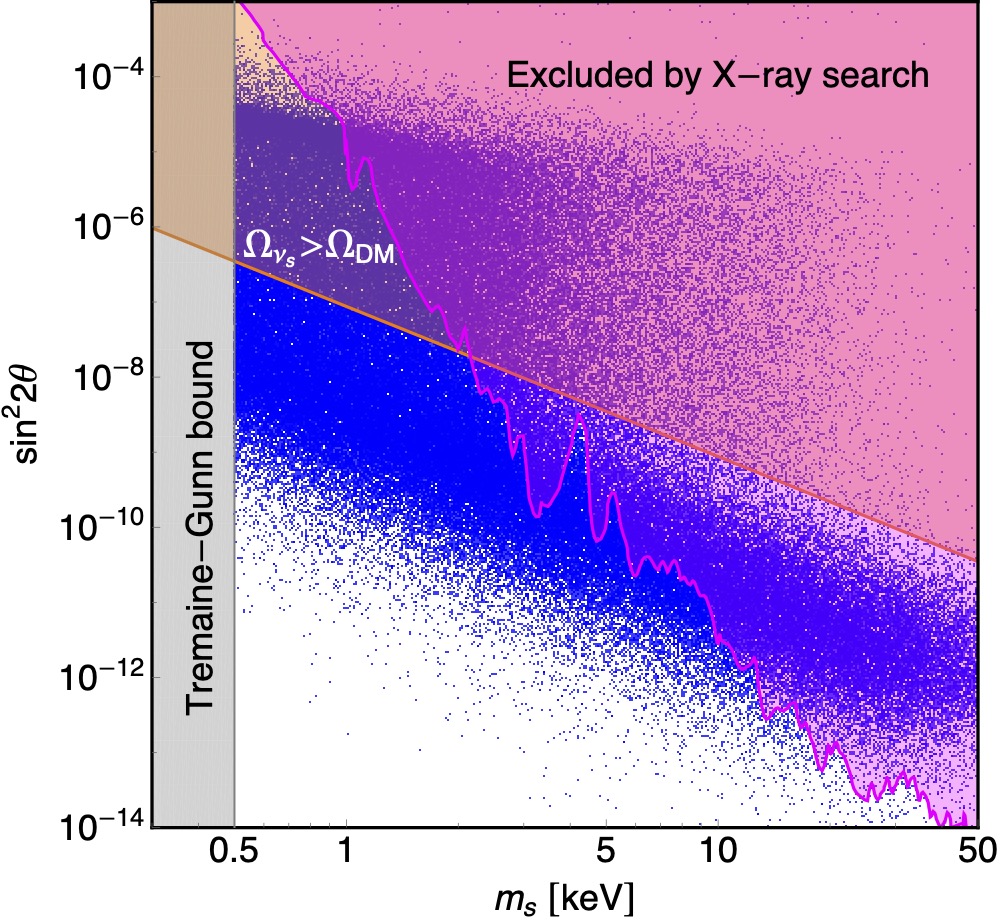}
\caption{The allowed parameter space shown in the $(m^{}_s,\sin^2 2\theta)$ plane. The light orange area is excluded by the overabundance of dark matter, while the pink area by the non-observation of the $X$-ray line, where the low-mass region is set mainly by Chandra~\cite{Horiuchi:2013noa} and XMM-Newton~\cite{Malyshev:2014xqa}, and the high-mass region by NuSTAR~\cite{Roach:2019ctw}. The Tremaine-Gunn bound~\cite{Tremaine:1979we} is also shown.}
\label{fig:asDM}
\end{figure}
%%%%%%%%%%%%%%%%%%%%%%%%%%%%%%%%%%%%%%%%%%%%%%%%%%%%%%%%%%%%%%%%%%%%%%%%%%

We now proceed with a numerical analysis of the parameter space by taking account of the constraints on WDM. We solve the active-sterile mixing from the full matrix $\mathcal{M}$ in Eq.~(\ref{eq:Mfull}). After diagonalization, the whole mass spectrum is given by $\widehat{\mathcal{M}}\equiv\mathrm{Diag}\{m^{}_1,m^{}_2,m^{}_3,m^{}_4,m^{}_5,m^{}_6,m^{}_7,m^{}_8\}$. From the generic feature of the mass spectrum in the ISS models as introduced in Sec.~\ref{subsec:wdm}, we identify that the light sterile mass is $m^{}_s\simeq m^{}_4$, and the heavy states form two pseudo-Dirac pairs with masses $m^{}_5 \simeq m^{}_6$ and $m^{}_7 \simeq m^{}_8$. To solve for the mixing and the mass spectrum, we need to restore two coefficient ratios from the three matrices: $M^{}_\mathrm{D},~ M^{}_\mathrm{S}, ~\mu$. To show it in a more explicit way, we examine all the three coefficients and set
\begin{align}
\Lambda  = 10^{2}~\mathrm{GeV},\quad
yv^{}_\mathrm{u}  \in [10^{-6},10^{-2}] ~\Lambda,\quad
g \in [10^{-4},10^{-2}] ~yv^{}_\mathrm{u}. \label{eq:spectrumPara}
%             (45)
\end{align}
Note that these coefficients do not affect oscillation phenomenology, as long as they fulfill the requirement that the approximation made in Eq.~(\ref{eq:mnu}) is valid, i.e., $g \ll yv^{}_\mathrm{u}, \Lambda$. As $\left(yv^{}_\mathrm{u} / \Lambda \right)^2$ characterizes the magnitude of the non-unitary effect, we require it less than $\mathcal{O}(10^{-4})$ to be in no conflict with existing experimental constraints. The value of $\Lambda$ is chosen only for illustrative purpose. The overall scale is set by the ratio of the neutrino mass-squared differences.

It can be inferred from Eq.~(\ref{eq:spectrumPara}) that the heavy-state masses can span a large range in the model. It is understandable since the inverse seesaw formula only requires $M^{}_\mathrm{D}/M^{}_\mathrm{S} \sim \mathcal{O}(10^{-2})$ even if we fix $\mu$ at keV. From a numerical scan, we select points with $m^{}_s \in [0.5,50]$ keV as the viable ones. For the viable points, we find $m^{}_5 \simeq m^{}_6$ and $m^{}_7 \simeq m^{}_8$ as expected. We plot the correlations of the active-sterile mixing, the lightest neutrino mass, two heavy-state masses with the scale ratio $yv^{}_\mathrm{u}/\Lambda$ in Fig.~\ref{fig:mms}. The ratio $g/(yv^{}_\mathrm{u})$ is insensitive to these quantities; thus, we do not show them here.

The constraints from the relic abundance, $X$-ray search, and the Tremaine-Gunn bound are shown in Fig.~\ref{fig:asDM}. From Fig.~\ref{fig:asDM} we see that given the light sterile mass in the desired region, the active-sterile mixing can span a wide range. Although a  large range has been excluded either by overabundance or $X$-ray search, there are still vast surviving points corresponding to an active-sterile mixing $\sin^22\theta$ from $10^{-14}$ to $10^{-7}$, depending on the sterile neutrino mass. From the upper-left plot in Fig.~\ref{fig:mms}, we see that this active-sterile mixing window requires $yv^{}_\mathrm{u}/\Lambda$ to be smaller than $10^{-4}$. As $yv^{}_\mathrm{u}/\Lambda$ characterizes $M^{}_\mathrm{D}/M^{}_\mathrm{S}$, and the non-unitary effect is measured by $M_\mathrm{D}^2/M_\mathrm{S}^2$, we conclude that the non-unitary effect is vanishingly small. Applying the same argument, we find
\begin{align}
m^{}_5 \simeq m^{}_6 \in [1, 10^{3}] ~\text{GeV},\quad m^{}_7 \simeq m^{}_8 \in [ 10, 10^{4}] ~\text{GeV} \;.
%             (46)
\end{align}
Given the above heavy state masses, one may expect their signatures at the high-energy hadron colliders. A detailed analysis will be carried out elsewhere. The relevant existing studies can be found in Refs.~\cite{Hirsch:2009ra,Das:2012ze,Bandyopadhyay:2012px,Das:2014jxa,Das:2015toa,Das:2018usr}.

By far we have not put the Lyman-$\alpha$ forest constraints for the reason that it is model-dependent as mentioned in Sec.~\ref{subsec:wdm}. The Lyman-$\alpha$ bounds on $m^{}_\mathrm{WDM}$ is derived by assuming a Fermi-Dirac distribution for WDM, which is neither the case for DW production nor the case for SF production. Light sterile neutrinos from DW production feature a momentum distribution that can be approximated by a rescaled Fermi-Dirac distribution, and also have nearly the same transfer function as the thermal relics. Therefore, the rescaling relation between $m_\mathrm{WDM}^{}$ and $m_s^{}$ is valid~\cite{Baur:2015jsy,Bozek:2015bdo}
\begin{align}
\frac{m^{}_s}{3.9 ~\mathrm{keV}} =\left( \frac{m^{}_{\mathrm{WDM}}}{\mathrm{keV}} \right)^{1.294} \left( \frac{0.25\times 0.7^2}{\Omega_\mathrm{DM}^{} h^2} \right)^{1/3}.
%             (47)
\end{align}
This rescaling is only approximate as the momentum distribution from the DW production mechanism differs from a rescaled Fermi-Dirac distribution. Applying the current Lyman-$\alpha$ bound on WDM in Eq.~(\ref{eq:Lya}), we get a lower bound on light sterile mass as
\begin{align}
m^{}_s \gtrsim \left(8.9-33.8 \right) ~\mathrm{keV}.
%             (48)
\end{align}
This lower bound clearly contradicts the $X$-ray bound, which limits the light sterile neutrino mass to be below $4$ keV.\footnote{It differs a little from direct observation of Fig.~\ref{fig:asDM}, as the dark matter abundance line is drawn with an approximation that all the flavor-dependent coefficients are equal to 0.5.} This combined result shows that DW-produced sterile neutrinos cannot be $100\%$ dark matter. This, however, does not rule out the sterile neutrino in this model as a viable dark matter candidate. In what follows, we present a few possible options.

First, if sterile neutrinos produced from the DW mechanism only contribute a fraction of total dark matter, the Lyman-$\alpha$ bounds can be relaxed, thus opening up viable parameter space.
Second, heavy states may have an impact as entropy dilution if they once dominate the energy density of the Universe. The inclusion of such an effect can also enlarge viable parameter space, as discussed in Refs.~\cite{Asaka:2006nq,Bezrukov:2009th,Abada:2014zra}.
Third, for SF production, the probability of active-sterile transition gets an enhancement when a lepton asymmetry dominates over the background potential and satisfies the resonant condition. The resulting momentum distribution is ``colder" than that of the DW mechanism, thus relaxes the constraints from the small-scale structure. As the momentum distribution is highly non-thermal, no mass-rescaling relation with the WDM can be derived. Hydrodynamical simulations in this case find that sterile neutrinos lighter than $7$ keV are inconsistent with the BOSS Lyman-$\alpha$ data~\cite{Baur:2017stq}. On the other hand, the resonance allows a small mixing for the sterile neutrino production, thus relaxes the $X$-ray bounds.

%Heavy states are consistent with a pre-existing lepton asymmetry at the order of ($10^{-6}$ -- $10^{-5}$).  One can find in Ref.~\cite{Eijima:2020shs} for possible ways to generate such a lepton asymmetry as required pre-existing.

\section{Conclusion}\label{sec:conclusion}

We construct simple models of neutrino masses and flavor mixing using the modular $S^{}_4$ as the flavor symmetry group. By introducing two right-handed neutrinos and three singlet fermions, we obtain light neutrino masses through the inverse seesaw mechanism, and we can have an intermediate-scale sterile neutrino which serves as a dark matter candidate. We maintain a minimal field content and utilize all the five irreducible representations in $S^{}_4$ in a natural way. By requiring that the modular forms having weights no larger than four and taking both light neutrino mass orderings into consideration, we consider twenty-four scenarios.

We perform a Bayesian model selection and find strong preference for the model \textbf{A1B2C2} with normal ordering. A detailed analysis shows that the model is in excellent agreement with observations from both the charged lepton and the neutrino sector. There are strong correlations of the model parameters, which reduce the effective degrees of freedom to six. The parameter correlation is explained in an analytical way. With nine model parameters, we get information about all the observables in both charged lepton and neutrino sector, i.e., three charged lepton masses, three light neutrino masses, three mixing angles, and three phases in the lepton mixing matrix, as well as the sterile neutrino spectrum and their mixing with the active neutrinos. From this viewpoint, the model is highly predictive and testable.

With a complete numerical analysis, we find that in the desired sterile neutrino mass range, a large range of the active-sterile mixing is still allowed. Stringent constraints come from the combination of the $X$-ray search and the Lyman-$\alpha$ forest data. Though being a $100$\% dark matter generated non-resonantly is ruled out, the sterile neutrino can still be a viable dark matter candidate with one of the following realizations: (i) contribute only a fraction of the total dark matter abundance; (ii) produced with an entropy dilution from the decays of the heavier sterile neutrino states; (iii) produced resonantly.

With an emphasis on the oscillation and dark matter phenomenology, we do not discuss possible collider signatures of the heavy sterile neutrinos. It is an interesting topic and will be studied elsewhere in the future.
To sum up, by natural and simple construction, we find one model among all the possibilities in excellent agreement with neutrino masses and mixing and it provides a viable dark matter candidate. The model is highly predictive and can be tested in future oscillation experiments and cosmological observations.

\section*{Acknowledgement}
This work was supported by the National Natural Science Foundation of China under Grants No. $11775232$ and No. $11835013$, and by the CAS Center for Excellence in Particle Physics.

\appendix
\section*{Appendices}
%\addcontentsline{toc}{section}{Appendices}
\renewcommand{\thesubsection}{\Alph{subsection}}

\subsection{Modular Group Theory}\label{apd:modular}
The modular group $\bar{\Gamma}$ is the group of linear fractional transformations
\begin{align}
\tau \rightarrow \gamma \tau= \frac{a\tau+b}{c\tau+d},~\text{with}~ a,b,c,d \in \mathbb{Z},~ ad-bc = 1,~ \mathrm{Im}\tau>0 \;,
\end{align}
which act faithfully on the upper-half complex plane. $\bar{\Gamma}$ is isomorphic to the projective special linear group PSL(2,$\mathbb{Z}$), which is defined as
\begin{align}
\left\{ \left(\begin{array}{cc}
a & b\\
c & d\\
\end{array}
\right):a,b,c,d \in \mathbb{Z}, ~ad-bc = 1\right\}/\{\pm1\}\; .
\end{align}
The PSL(2,$\mathbb{Z}$) group has two generators
\begin{align}
S=\left(\begin{array}{cc}
0 & 1\\
-1 & 0\\
\end{array}
\right), \quad
T=\left(\begin{array}{cc}
1 & 1\\
0 & 1\\
\end{array}
\right)\;.
\end{align}
The generators satisfy a minimal set of relation $S^2=(ST)^3=I$. Applying $S$, $T$ to $\tau$, one finds that the two generating transformations for $\bar{\Gamma}$ are
\begin{align}
\tau \xrightarrow[]{S} -\frac{1}{\tau},\quad \tau \xrightarrow[]{T} \tau + 1\;.
\end{align}
The principal congruence subgroups $\Gamma(N), N=1,2,3,...$ are defined as
\begin{align}
\Gamma(N) =\left\{\left(\begin{array}{cc}
a & b\\
c & d\\
\end{array}
\right)\in \mathrm{SL}(2,\mathbb{Z}),~
\left(\begin{array}{cc}
a & b\\
c & d\\
\end{array}
\right)=\left(\begin{array}{cc}
1 & 0\\
0 & 1\\
\end{array}
\right) (\text{mod}~ N)\right\}\;.
\end{align}
$\Gamma(N)$ are infinite normal subgroups of the special linear group SL$(2,\mathbb{Z})=\Gamma(1)=\Gamma$, which is the group of 2$\times$2 matrices with integer elements and determinant $1$.
With
\begin{align}
N&=1,2,\quad\bar{\Gamma}(N)\equiv\Gamma(N)/\{I,\pm I\}\;;\\
N&>2,\quad \bar{\Gamma}(N)\equiv\Gamma(N)\;,
\end{align}
one can introduce the finite modular groups as the quotient groups $\Gamma^{}_N\equiv \bar{\Gamma}/\bar{\Gamma}(N)$. For $N\leqslant 5$, $\Gamma^{}_N$ are isomorphic to permutation groups, i.e., $\Gamma^{}_2 \simeq S^{}_3$, $\Gamma^{}_3 \simeq A^{}_4$, $\Gamma^{}_4 \simeq S^{}_4$, and $\Gamma^{}_5 \simeq A^{}_5$.

In modular-invariant supersymmetric theories (SUSY), e.g., $\mathcal{N}=1$ global SUSY, the action is
\begin{align}
\mathcal{S} =\int d^4 x d^2 \theta d^2 \bar{\theta} K(\chi_i^{}, \bar{\chi}_i^{}) +\int d^4x d^2\theta W(\chi_i^{}) + \mathrm{h.c. }\;,
\end{align}
where $K$ is the K\"{a}hler potential, $W$ is the superpotential and $\chi_i^{}$ are the chiral superfields. Requiring the action being invariant under $\Gamma_N^{}$ leads to the following transformation for the chiral superfields
\begin{align}
\chi_i^{} \rightarrow (c \tau + d)^{-k_i^{}} \rho_i^{}\left(\gamma \right) \chi_i^{}\;,
\end{align}
where $\rho_i^{}$ are the unitary representation matrices and $k_i^{}$ are modular weights carried by superfields $\chi_i^{}$.
The invariance of the action requires the superpotential being invariant while the K\"{a}hler potential is invariant up to a K\"{a}hler transformation,
\begin{align}
W(\chi^{}_i) &\rightarrow W(\chi^{}_i)\;,\\
K(\chi^{}_i, \bar{\chi}^{}_i) &\rightarrow K(\chi^{}_i, \bar{\chi}^{}_i) + f(\chi^{}_i) + f(\bar{\chi}^{}_i)\;.
\end{align}
It is more relevant to focus on the superpotential, which can be expressed as
\begin{align}
W =\sum_n \sum_{\{ i_1,...,i_n\}} \left( Y^{}_{\{ i_1,...,i_n\}} (\tau) \chi^{}_{i_1},...,\chi^{}_{i_n} \right)_\textbf{1}\;.
\end{align}
The invariance of the superpotential requires the coefficient functions $Y(\tau)$ transforming under $\Gamma^{}_N$ as
\begin{align}
Y(\tau) \rightarrow Y(\gamma \tau)=(c \tau + d)^{k^{}_Y}_{} \rho^{}_Y(\gamma) Y(\tau)\;.
\end{align}
To make the superpotential invariant, the modular weights and the representations should satisfy
\begin{align}
&k^{}_Y =k^{}_{i_1} +...+k^{}_{i_n}\;,\\
&\rho^{}_Y \otimes \rho^{}_{I_1} \otimes ...\supset \textbf{1}\;.
\end{align}

\subsection{Modular Group $S^{}_4$}\label{apd:S4}
The $S^{}_4$ group is made from permutations of four objects. It has five irreducible representations: $\bf{1}, \bf{1^\prime}, \bf{2}, \bf{3}$ and $\bf{3^\prime}$, and two generators satisfying the following condition:
\begin{align}
S^2 = T^4 = (ST)^3 = I\; .
\end{align}
Working in a symmetric basis, we collect the expressions of the generators, the Clebsch-Gordan coefficients, the $q$-expansion of the basis, and the modular-form multiplets of the low weights here for reader's convenience. These results are taken from Ref.~\cite{Novichkov:2019sqv}.

The generators in the five irreducible representations are all symmetric matrices, i.e.,
\begin{align}
&\textbf{1}: S=1, \quad T=1\;, \\
&\textbf{1}^\prime: S=-1, \quad T=-1\;,\\
&\textbf{2}: S=\frac{1}{2}
\left(\begin{array}{cc}
-1 & \sqrt{3}\\
\sqrt{3} & 1 \\
\end{array}\right)\; , \quad T=
\left(\begin{array}{cc}
1 & 0\\
0 & -1 \\
\end{array}\right)\;,\\
&\textbf{3}: S=-\frac{1}{2}
\left(\begin{array}{ccc}
0 & \sqrt{2} & \sqrt{2}\\
\sqrt{2} & -1 & 1 \\
\sqrt{2} & 1 & -1 \\
\end{array}\right)\; , \quad T= -
\left(\begin{array}{ccc}
1 & 0 & 0\\
0 & i & 0 \\
0 & 0 & -i \\
\end{array}\right)\;,\\
&\textbf{3}^\prime: S=\frac{1}{2}
\left(\begin{array}{ccc}
0 & \sqrt{2} & \sqrt{2}\\
\sqrt{2} & -1 & 1 \\
\sqrt{2} & 1 & -1 \\
\end{array}\right)\; , \quad T=
\left(\begin{array}{ccc}
1 & 0 & 0\\
0 & i & 0 \\
0 & 0 & -i \\
\end{array}\right)\;.
\end{align}

From the generator matrices, one can derive the Clebsch-Gordan coefficients of tensor products of two multiplets $\alpha$ and $\beta$ as follows.
\begin{align}
&\textbf{1}^\prime \otimes \textbf{1}^\prime =\textbf{1}\;,  \quad \textbf{1}\sim
\alpha_1 \beta_1\;,\\
&\textbf{1}^\prime \otimes \textbf{2} =\textbf{2}\;, \quad \textbf{2} \sim
\left( \begin{array}{c}
-\alpha_1 \beta_2 \\
\alpha_1 \beta_1 \\
\end{array} \right)\;,\\
&\textbf{1}^\prime \otimes \textbf{3} = \textbf{3}^\prime\;, \quad \textbf{3}^\prime \sim
\left(\begin{array}{c}
\alpha_1 \beta_1 \\
\alpha_1 \beta_2 \\
\alpha_1 \beta_3 \\
\end{array}
\right)\;,\\
&\textbf{1}^\prime \otimes \textbf{3}^\prime = \textbf{3}\;, \quad \textbf{3} \sim
\left(\begin{array}{c}
\alpha_1 \beta_1 \\
\alpha_1 \beta_2 \\
\alpha_1 \beta_3 \\
\end{array}
\right)\;,\\
\nonumber\\
&\textbf{2} \otimes \textbf{2} = \textbf{1} \oplus \textbf{1}^\prime \oplus \textbf{2} \\
&\quad \quad \textbf{1} \sim \alpha_{1} \beta_1 + \alpha_2 \beta_2 \;, \quad
\textbf{1}^\prime \sim \alpha_1 \beta_2 - \alpha_2 \beta_1\;, \quad
\textbf{2} \sim
\left(\begin{array}{c}
\alpha_2 \beta_2 - \alpha_1 \beta_1 \\
\alpha_1 \beta_2 + \alpha_2 \beta_1 \\
\end{array}
\right) \;,\nonumber\\
\nonumber\\
&\textbf{2} \otimes \textbf{3} = \textbf{3} \oplus \textbf{3}^\prime  \\
&\quad \quad \textbf{3} \sim
\left(\begin{array}{c}
 \alpha_1 \beta_1 \\
\frac{\sqrt{3}}{2}\alpha_2 \beta_3 - \frac{1}{2}\alpha_1 \beta_2 \\
\frac{\sqrt{3}}{2}\alpha_2 \beta_2 - \frac{1}{2}\alpha_1 \beta_3 \\
\end{array}
\right)\;, \quad
\textbf{3}^\prime \sim
\left(\begin{array}{c}
 -\alpha_2 \beta_1 \\
\frac{\sqrt{3}}{2}\alpha_1 \beta_3 + \frac{1}{2}\alpha_2 \beta_2 \\
\frac{\sqrt{3}}{2}\alpha_1 \beta_2 + \frac{1}{2}\alpha_2 \beta_3 \\
\end{array}
\right)\;, \nonumber\\
\nonumber\\
&\textbf{2} \otimes \textbf{3}^\prime = \textbf{3} \oplus \textbf{3}^\prime  \\
&\quad \quad \textbf{3} \sim
\left(\begin{array}{c}
 -\alpha_2 \beta_1 \\
\frac{\sqrt{3}}{2}\alpha_1 \beta_3 + \frac{1}{2}\alpha_2 \beta_2 \\
\frac{\sqrt{3}}{2}\alpha_1 \beta_2 + \frac{1}{2}\alpha_2 \beta_3 \\
\end{array}
\right)\;, \quad
\textbf{3}^\prime \sim
\left(\begin{array}{c}
 \alpha_1 \beta_1 \\
\frac{\sqrt{3}}{2}\alpha_2 \beta_3 - \frac{1}{2}\alpha_1 \beta_2 \\
\frac{\sqrt{3}}{2}\alpha_2 \beta_2 - \frac{1}{2}\alpha_1 \beta_3 \\
\end{array}
\right)\;, \nonumber\\
\nonumber\\
&\textbf{3} \otimes \textbf{3} = \textbf{3}^\prime \otimes \textbf{3}^\prime = \textbf{1} \oplus \textbf{2} \oplus \textbf{3} \oplus \textbf{3}^\prime  \\
&\quad \quad \textbf{1} \sim \alpha_1 \beta_1 + \alpha_2 \beta_3 + \alpha_3 \beta_2\;, \quad
\textbf{2} \sim
\left(\begin{array}{c}
\alpha_1 \beta_1-\frac{1}{2}\left(\alpha_2 \beta_3 + \alpha_3 \beta_2 \right)\\
\frac{\sqrt{3}}{2} \left(\alpha_2 \beta_2 + \alpha_3 \beta_3 \right)\\
\end{array}
\right) \;,\nonumber\\
&\quad \quad\textbf{3} \sim
\left(\begin{array}{c}
\alpha_3 \beta_3 - \alpha_2 \beta_2 \\
\alpha_1 \beta_3 + \alpha_3 \beta_1 \\
-\alpha_1 \beta_2 - \alpha_2 \beta_1 \\
\end{array}
\right)\;, \quad
\textbf{3}^\prime \sim
\left(\begin{array}{c}
 \alpha_3 \beta_2 - \alpha_2 \beta_3 \\
\alpha_2 \beta_1 - \alpha_1 \beta_2 \\
\alpha_1 \beta_3 - \alpha_3 \beta_1 \\
\end{array}
\right)\;, \nonumber\\
\nonumber\\
&\textbf{3} \otimes \textbf{3}^\prime =  \textbf{1}^\prime \oplus \textbf{2} \oplus \textbf{3} \oplus \textbf{3}^\prime  \\
&\quad \quad\textbf{1}^\prime \sim \alpha_1 \beta_1 + \alpha_2 \beta_3 + \alpha_3 \beta_2\;, \quad
\textbf{2} \sim
\left(\begin{array}{c}
\frac{\sqrt{3}}{2}\left(\alpha_2 \beta_2 + \alpha_3 \beta_3 \right)\\
-\alpha_1 \beta_1 +\frac{1}{2} \left(\alpha_2 \beta_3 + \alpha_3 \beta_2 \right)\\
\end{array}
\right)\;, \nonumber\\
&\quad \quad \textbf{3} \sim
\left(\begin{array}{c}
\alpha_3 \beta_2 - \alpha_2 \beta_3 \\
\alpha_2 \beta_1 - \alpha_1 \beta_2 \\
\alpha_1 \beta_3 - \alpha_3 \beta_1 \\
\end{array}
\right)\;, \quad
\textbf{3}^\prime \sim
\left(\begin{array}{c}
 \alpha_3 \beta_3 - \alpha_2 \beta_2 \\
\alpha_1 \beta_3 + \alpha_3 \beta_1 \\
-\alpha_1 \beta_2 - \alpha_2 \beta_1 \\
\end{array}
\right)\;. \nonumber
\end{align}

The basis in the space of the lowest-weight modular forms can be written in $q$-expansion as
\begin{align}
Y^{}_1 &= -3 \pi \left( \frac{1}{8} + 3q + 3q^2 +12 q^3 + 3 q^4 + 18 q^5 +12 q^6 + 24 q^7 +3 q^8 +39 q^9 \right)\; ;\\
Y^{}_2 &= 3\sqrt{3} \pi q^{1/2} \left( 1 + 4q + 6q^2 + 8 q^3 + 13 q^4 + 12 q^5 + 14 q^6 + 24 q^7 + 18 q^8 + 20 q^9\right)\; ;\\
Y^{}_3 &=  \pi \left( \frac{1}{4} - 2q + 6q^2 - 8 q^3 + 6 q^4 - 12 q^5 + 24 q^6 - 16 q^7 + 6 q^8 - 26 q^9 + 38 q^{10}\right)\; ;\\
Y^{}_4 &= -\sqrt{2} \pi q^{1/4} \left( 1 + 6q + 13q^2 + 14 q^3 + 18 q^4 + 32 q^5 + 31 q^6 + 30 q^7 + 48 q^8 + 38 q^9\right)\; ;\\
Y^{}_5 &= -4\sqrt{2} \pi q^{3/4} \left( 1 + 2q + 3q^2 + 6 q^3 + 5 q^4 + 6 q^5 + 10 q^6 + 8 q^7 + 12 q^8 + 14 q^9\right)\; ,
\end{align}
where $q\equiv e^{\mathrm{i} 2\pi \tau}$. The modular-form multiplets of the lowest weight are
\begin{align}
Y^{}_\textbf{2}=\left(
\begin{array}{c}
Y^{}_1\\
Y^{}_2\\
\end{array}\right), \quad
Y^{}_{\textbf{3}^\prime} = \left(
\begin{array}{c}
Y^{}_3\\
Y^{}_4\\
Y^{}_5\\
\end{array}\right)\; .
\end{align}
At weight four, the modular-form multiplets are
\begin{align}
Y_\textbf{1}^{(4)} &= Y_1^2+ Y_2^2, \quad
Y_\textbf{2}^{(4)} = \left(
\begin{array}{c}
Y_2^2 -Y_1^2 \\
2 Y^{}_1 Y^{}_2 \\
\end{array} \right),\\
Y_\textbf{3}^{(4)} &=\left(
\begin{array}{c}
-2Y^{}_2  Y^{}_3 \\
\sqrt{3} Y^{}_1 Y^{}_5 + Y^{}_2 Y^{}_4 \\
\sqrt{3} Y^{}_1 Y^{}_4 + Y^{}_2 Y^{}_5 \\
\end{array} \right), \quad
Y_{\textbf{3}^\prime}^{(4)} =\left(
\begin{array}{c}
2Y^{}_1  Y^{}_3 \\
\sqrt{3} Y^{}_2 Y^{}_5 - Y^{}_1 Y^{}_4 \\
\sqrt{3} Y^{}_2 Y^{}_4 - Y^{}_1 Y^{}_5 \\
\end{array} \right)\; ,
\end{align}
where we use superscript ``$(4)$" to explicitly indicate the modular weight.

\subsection{Block Diagonalization in ISS(2,3) Models}\label{apd:block}
Block diagonalization in ISS(2,3) models is distinct from that in the $n^{}_{N_\mathrm{R}} = n^{}_{S_\mathrm{L}}$  ISS models, mainly due to the rectangular matrices $M^{}_\mathrm{D}$ and $M^{}_{S}$ in the former cases. We firstly review the block diagonalization procedure in type-I seesaw~\cite{Ibarra:2010xw,Merle:2013gea}, which is useful in later discussions.

Consider the full neutrino mass matrix in type-I seesaw with three left-handed and $k$ right-handed neutrinos
\begin{align}
\left(\begin{array}{cc}
{\bf{0}}^{}_{3\times 3} & \left[ M^{}_\mathrm{D} \right]^{}_{3\times k}\\
\left[M_\mathrm{D}^\mathrm{T}\right]^{}_{k\times 3} & \left[ M^{}_N \right]^{}_{k\times k}\\
\end{array}
\right)\; ,
%           (87)
\end{align}
which can be block-diagonalized by a unitary matrix like
\begin{align}
\Omega^\mathrm{T}_{}
\left(\begin{array}{cc}
{\bf{0}}^{}_{3\times 3} & \left[ M^{}_\mathrm{D} \right]^{}_{3\times k}\\
\left[M_\mathrm{D}^\mathrm{T}\right]^{}_{k\times 3} & \left[ M^{}_N \right]^{}_{k\times k}\\
\end{array}
\right) \Omega
=\left(\begin{array}{cc}
\left[ U \widehat{M}_\nu U^\mathrm{T}_{}\right]^{}_{3\times 3} & {\bf{0}}^{}_{3\times k} \\
{\bf{0}}^{}_{k\times 3}  & \left[ V \widehat{M}_N V_{}^\mathrm{T} \right]^{}_{k \times k}\\
\end{array}
\right)\; ,
%           (88)
\end{align}
where $U$ and $V$ are unitary matrices and $M^{}_\nu=U \widehat{M}^{}_\nu U_{}^\mathrm{T}$.
%~M^{}_N=V \widehat{M}^{}_N V_{}^\mathrm{T}$.
With
\begin{align}
\Omega=\mathrm{exp} \left(\begin{array}{cc}
\bf{0} & R\\
-R^\dagger_{} & \bf{0}\\
\end{array}\right)
= \left(\begin{array}{cc}
1-\frac{1}{2} RR^\dagger_{} & R\\
-R^\dagger_{} & 1-\frac{1}{2} R^\dagger_{} R\\
\end{array}\right)
+\mathcal{O}(R^3)\; ,
%           (89)
\end{align}
we find
\begin{align}
U \widehat{M}_\nu^{} U^\mathrm{T}_{} &= -R^*_{} M^{}_N R^\dagger_{}, \label{eq:typeI_raw}\\
%           (90)
V \widehat{M}_N^{} V^\mathrm{T}_{}&=  M^{}_N +\frac{1}{2} M^{}_N R^\dagger_{} R+ \frac{1}{2} R^\mathrm{T}_{}R^*_{}M^{}_N\; .
%           (91)
\end{align}
The vanishing non-diagonal entities give
\begin{align}
R^*_{}= M^{}_\mathrm{D} M_N^{-1}\; .
%           (92)
\end{align}
Substituting $R$ in Eq.~(\ref{eq:typeI_raw}), we find the type-I seesaw formula $M^{}_\nu= - M^{}_\mathrm{D} M_N^{-1} M_\mathrm{D}^\mathrm{T}$.

A full mass matrix in ISS models with $n^{}_{N_\mathrm{R}} = n^{}_{S_\mathrm{L}}$ is
\begin{align}
\left(
\begin{array}{ccc}
\bf{0} & m^{}_\mathrm{D} & \bf{0}\\
m_\mathrm{D}^\mathrm{T} & \bf{0}   & M_\mathrm{S}^\mathrm{T}\\
\bf{0} & M^{}_\mathrm{S} & \mu\\
\end{array}
\right)
= \left(
\begin{array}{cc}
\bf{0} & M^{}_\mathrm{D}\\
M_\mathrm{D}^\mathrm{T} & M^{}_N\\
\end{array}
\right)\; ,
\end{align}
where we define
\begin{align}
M^{}_\mathrm{D} =\left( \begin{array}{cc}
m^{}_\mathrm{D} & \bf{0}\\
\end{array}\right),\quad
M^{}_N = \left(\begin{array}{cc}
\bf{0} & M_\mathrm{S}^\mathrm{T}\\
M^{}_\mathrm{S} & \mu\\
\end{array}\right)\; .
%      (94)
\end{align}
With
\begin{align}
R^*_{}=M^{}_\mathrm{D} M_N^*
= \begin{array}{cc}
-M_\mathrm{S}^{-1} \mu (M_\mathrm{S}^{-1})^\mathrm{T} m_\mathrm{D}^\mathrm{T} & (M_\mathrm{S}^{-1})^\mathrm{T}m_\mathrm{D}^\mathrm{T}\\
\end{array}\; ,
%  (95)
\end{align}
we get
\begin{align}
M^{}_\nu = -R^*_{} M^{}_N R^\dagger_{}
= -m^{}_\mathrm{D} M_\mathrm{S}^{-1} \mu (M_\mathrm{S}^{-1})^\mathrm{T} m_\mathrm{D}^\mathrm{T}\; . \label{eq:iss}
%.      (96)
\end{align}

Directly applying Eq.~(\ref{eq:iss}) to the $n^{}_{N_\mathrm{R}} \neq n^{}_{S_\mathrm{L}}$ case is not possible as $M^{}_\mathrm{S}$ is a rectangular matrix which has no inverse, but the procedure works in the same way. Considering the full mass matrix in Eq.~(\ref{eq:Mfull}) in our model, with
\begin{align}
M_\mathrm{D}^\prime = \left(\begin{array}{cc}
\left[ M^{}_\mathrm{D} \right]^{}_{3\times2}& \textbf{0}^{}_{3\times 3}\\
\end{array}
\right)\; , \quad
M_N^\prime = \left(\begin{array}{cc}
\textbf{0}^{}_{2\times 2} & \left[M_\mathrm{S}^\mathrm{T}\right]^{}_{2\times3} \\
%.
\left[M^{}_\mathrm{S} \right]^{}_{3\times2}     & \left[ \mu \right]^{}_{3\times 3}\\
\end{array}
\right)\; ,
%.       (97)
\label{eq:mnp}
\end{align}
we can express the light neutrino mass matrix in the same form as that of type-I seesaw, i.e., $M^{}_\nu= - M_\mathrm{D}^\prime \left(M_N^\prime\right)^{-1}_{} M_\mathrm{D}^{\prime \mathrm{T}}$. All we need is to find the inverse matrix of $M_N^\prime$, which does exist
\begin{align}
M_N^{\prime -1} = \left(\begin{array}{cc}
-\left[\left(M_\mathrm{S}^\mathrm{T} M^{}_\mathrm{S} M_\mathrm{S}^\mathrm{T} M^{}_\mathrm{S}\right)^{-1} M_\mathrm{S}^\mathrm{T} \mu M^{}_\mathrm{S} \right]^{}_{2\times 2} & \left[\left(M_\mathrm{S}^\mathrm{T} M^{}_\mathrm{S} \right)^{-1} M_\mathrm{S}^\mathrm{T}\right]^{}_{2\times 3}\\
\left[\left(M^{}_\mathrm{S} M_\mathrm{S}^\mathrm{T} \right)^{-1} M^{}_\mathrm{S}\right]^{}_{3\times 2} & \textbf{0}^{}_{3\times 3}\\
\end{array}
\right)^{}_{5\times 5}\; .
%           (98)
\label{eq:mni}
\end{align}
Substituting $M_\mathrm{D}^\prime$ and $M_N^{\prime -1}$ in Eq.~(\ref{eq:mnp}) and Eq.~(\ref{eq:mni}) into $M_\nu= - M_\mathrm{D}^\prime \left(M_N^\prime\right)^{-1}_{} M_\mathrm{D}^{\prime\mathrm{T}}$, we get the expression in Eq.~(\ref{eq:mnu}).

\end{document}